%% file: main.tex
\documentclass[conference]{IEEEtran}
\IEEEoverridecommandlockouts
\usepackage{cite}
\usepackage{amsmath,amssymb,amsfonts}
\usepackage{algorithmic}
\usepackage{graphicx}
\usepackage{textcomp}
\usepackage{xcolor}

\usepackage{mathptmx} 
\usepackage{cite}
\usepackage{amsmath,amssymb,amsfonts}
\usepackage{algorithmic}
\usepackage{graphicx}
\usepackage{textcomp}
\usepackage{xcolor}
\usepackage[hyphens]{url}
\usepackage{multirow}
\usepackage{caption}
\usepackage{subcaption}

\newcommand{\name}{{ALCOP}}
\newcommand{\hpcarevise}[1]{{#1}}

\def\BibTeX{{\rm B\kern-.05em{\sc i\kern-.025em b}\kern-.08em
    T\kern-.1667em\lower.7ex\hbox{E}\kern-.125emX}}

\begin{document}

\title{ALCOP: \underline{A}utomatic \underline{L}oad-\underline{CO}mpute \underline{P}ipelining in Deep Learning Compiler for AI-GPUs}

\author{\IEEEauthorblockN{Guyue Huang$^{1*}$, Yang Bai$^2$, Liu Liu$^3$, Yuke Wang$^1$, Bei Yu$^2$, Yufei Ding$^1$, Yuan Xie$^{1,4}$}
\IEEEauthorblockA{$^1$UC Santa Barbara, 
$^2$Chinese University of Hong Kong,
$^3$Rensselaer Polytechnic Institute, 
$^4$Alibaba DAMO Academy\\
*guyue@ucsb.edu}}

\maketitle

\begin{abstract}
\input{0_v3abstract}
\end{abstract}

\input{1_v5introduction}

\input{2_v2background}
\input{3_auto_scheduling}
\input{4_v2program_transformation}
\input{5_v2perfmodel}
\input{6_evaluation}
\input{7_v2relatedwork}
\input{8_conclusion}

\bibliographystyle{IEEEtran}
\bibliography{refs}
\end{document}

%% file: 0_v3abstract.tex
Pipelining between data loading and computation is a critical tensor program optimization for GPUs. In order to unleash the high performance of latest GPUs, we must perform a synergetic optimization of multi-stage pipelining across the multi-level buffer hierarchy of GPU. Existing frameworks rely on hand-written libraries such as cuBLAS to perform pipelining optimization, which is inextensible to new operators and un-composable with prior tensor compiler optimizations. This paper presents {\name}, the first framework that is compiler-native and fully supports multi-stage multi-level pipelining. {\name} overcomes three critical obstacles in generating code for pipelining: detection of pipelining-applicable buffers, program transformation for multi-level multi-stage pipelining, and efficient schedule parameter search by incorporating static analysis. Experiments show that {\name} can generate programs with 1.23$\times$ speedup on average (up to 1.73$\times$) over vanilla TVM. On end-to-end models, {\name} can improve upon TVM by up to 1.18$\times$, and XLA by up to 1.64$\times$. Besides, our performance model significantly improves the efficiency of the schedule tuning process and can find schedules with 99\% of the performance given by exhaustive search while costing $40\times$ fewer trials.

%% file: 1_v5introduction.tex
\begin{figure*}[!t]
    \centering
    \begin{subfigure}{0.55\textwidth}
    \includegraphics[width=\linewidth]{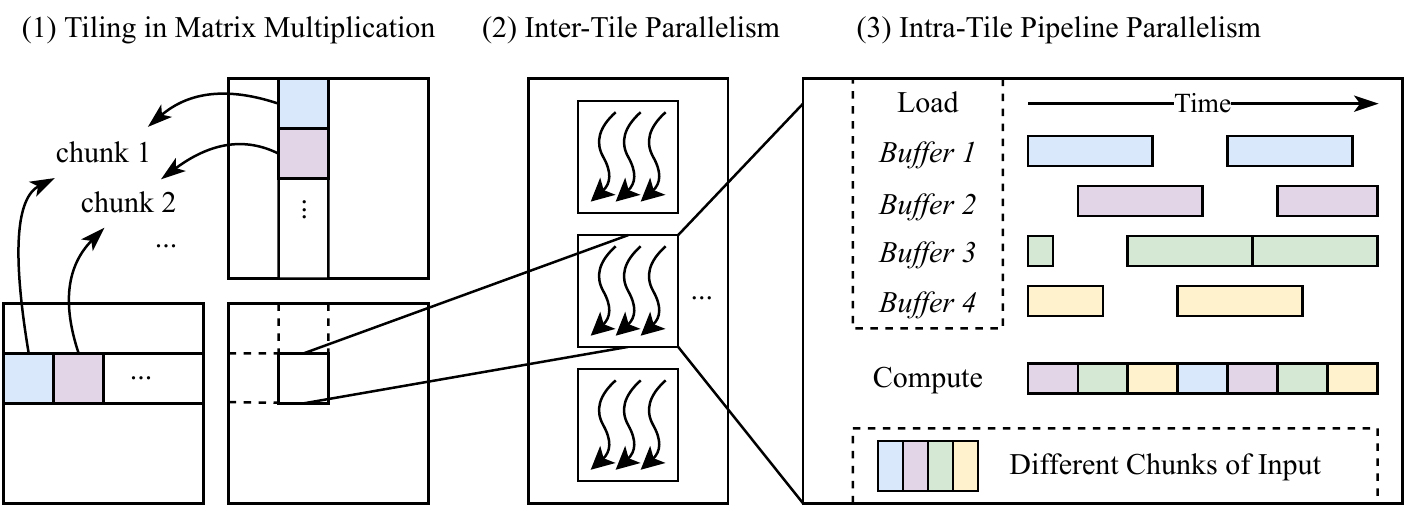}
    \caption{Concept of tiling, inter-tile parallelism and pipeline parallelism.}\label{fig:motivation:concept}
    \end{subfigure}
    \hfill
    \begin{subfigure}{0.43\textwidth}
    \includegraphics[width=\linewidth]{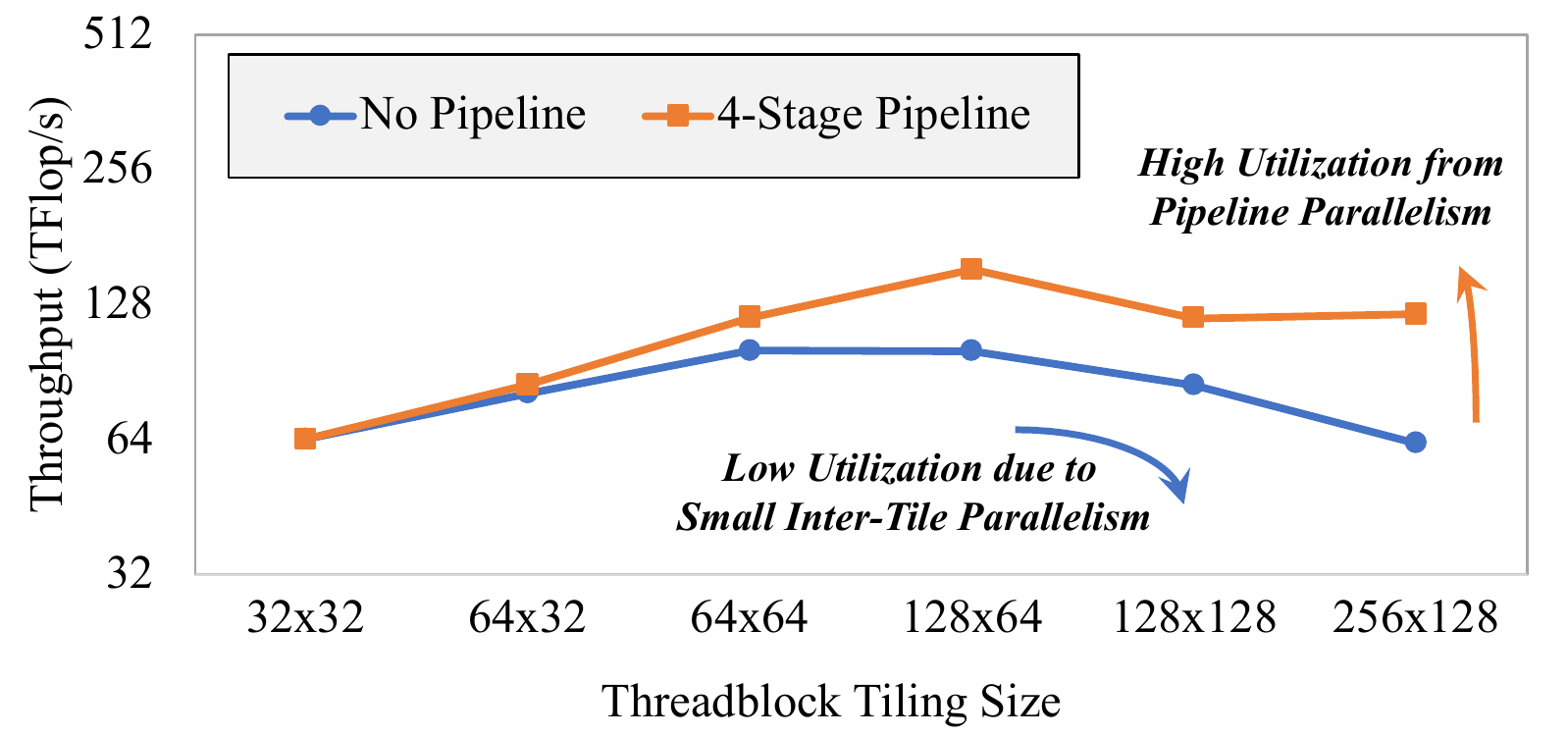}
    \caption{Motivating example: performance of a $2048\times2048\times2048$ matrix-multiplication \hpcarevise{tested on NVIDIA A100} with different tiling and pipelining choices.}\label{fig:motivation:example}
    \end{subfigure}
    \caption{Motivation of automatic pipelining. 
    (a-3) explains the concepts of pipelining, which is overlapping data loading with computation. (b) gives a motivating example. With tiling only, the performance is always sub-optimal. Pipelining unleashes intra-tile parallelism and increases the performance under large tiling. }
    \label{fig:motivation}
\end{figure*}

\section{Introduction}\label{sec:introduction}

\begin{figure}[tb!]
    \centering
    \begin{subfigure}{\linewidth}
    \centering
    \includegraphics[width=0.85\linewidth]{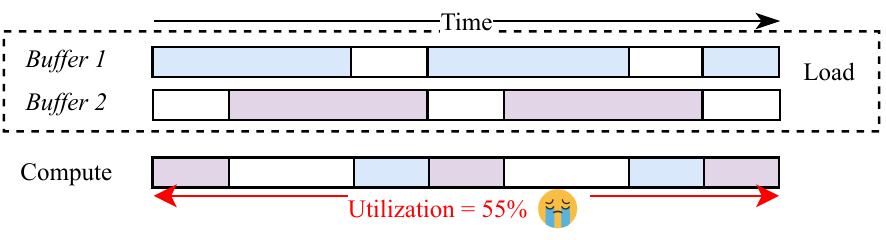}
    \caption{Two-stage pipelining}
    \label{fig:pipeline_multistage_1}
    \end{subfigure}
    \begin{subfigure}{\linewidth}
    \centering
    \includegraphics[width=0.85\linewidth]{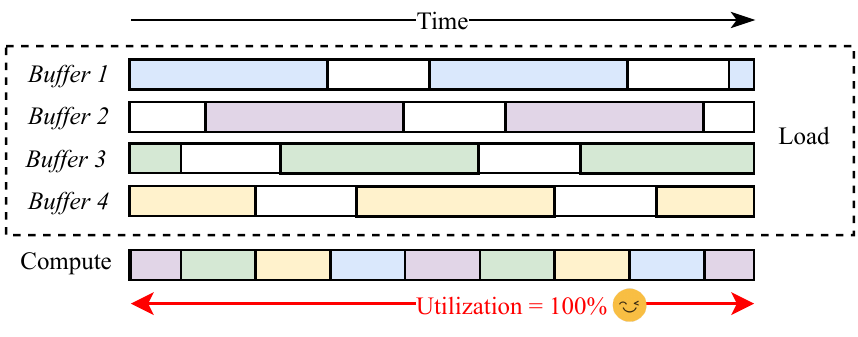}
    \caption{Multi-stage pipelining (four-stage here)}
    \label{fig:pipeline_multistage_2}
    \end{subfigure}
    \caption{Concept of multi-stage pipelining. (a) two-stage pipelining (or called double-buffering) is not enough to hide the data loading latency. 
    (b) Four-stage pipelining can hide the data loading latency and achieve full utilization of the computing units. {\name} supports multi-stage pipelining.}
    \label{fig:pipeline_multistage}
\end{figure}

\begin{figure}[tb!]
    \centering
    \begin{subfigure}{\linewidth}
    \centering
    \includegraphics[width=0.75\linewidth]{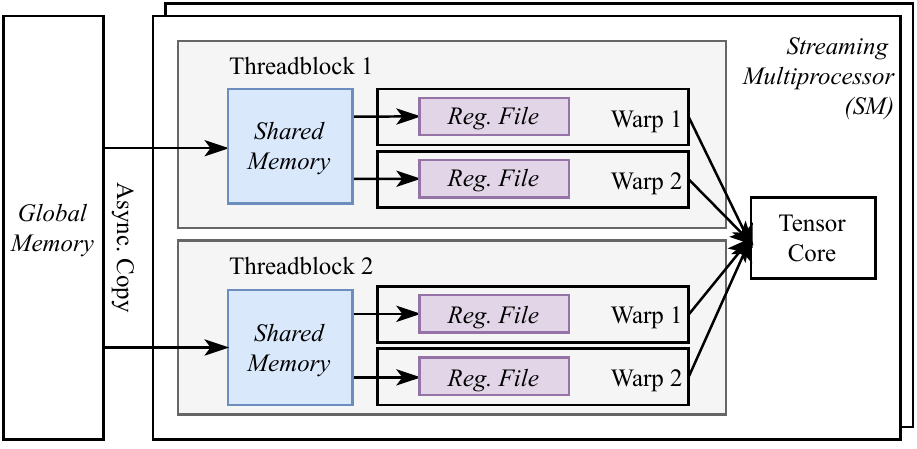}
    \caption{GPU memory hierarchy.}
    \label{fig:pipeline_multilevel:1}
    \end{subfigure}
    \begin{subfigure}{\linewidth}
    \centering
    \includegraphics[width=0.75\linewidth]{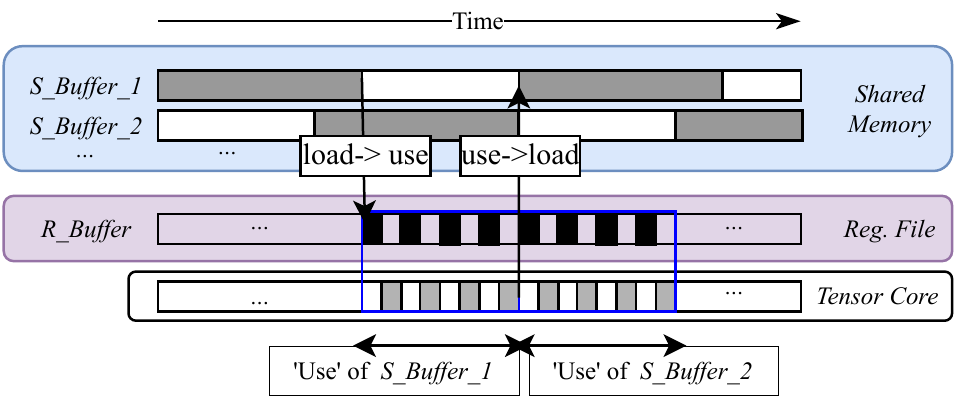}
    \caption{Single-level pipelining.}
    \label{fig:pipeline_multilevel:2}
    \end{subfigure}
    \begin{subfigure}{\linewidth}
    \centering
    \includegraphics[width=0.75\linewidth]{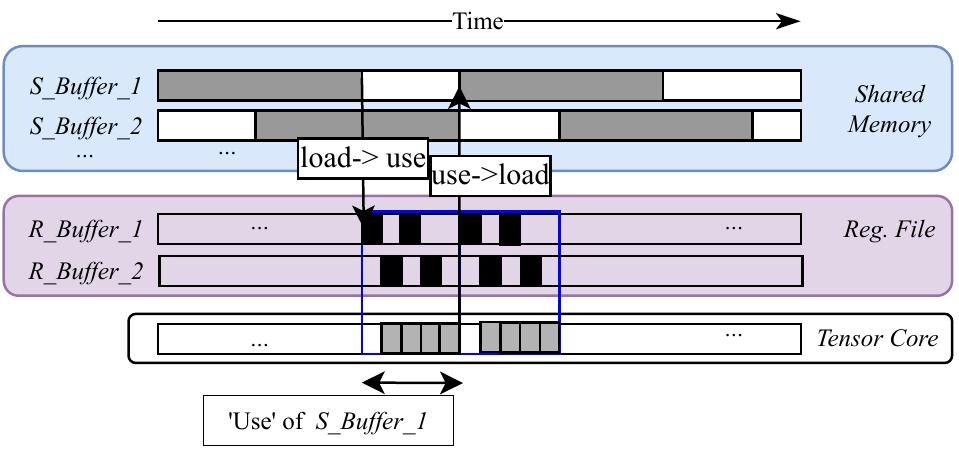}
    \caption{Multi-level pipelining without inner-pipeline fusion.}
    \label{fig:pipeline_multilevel:3}
    \end{subfigure}
    \begin{subfigure}{\linewidth}
    \centering
    \includegraphics[width=0.75\linewidth]{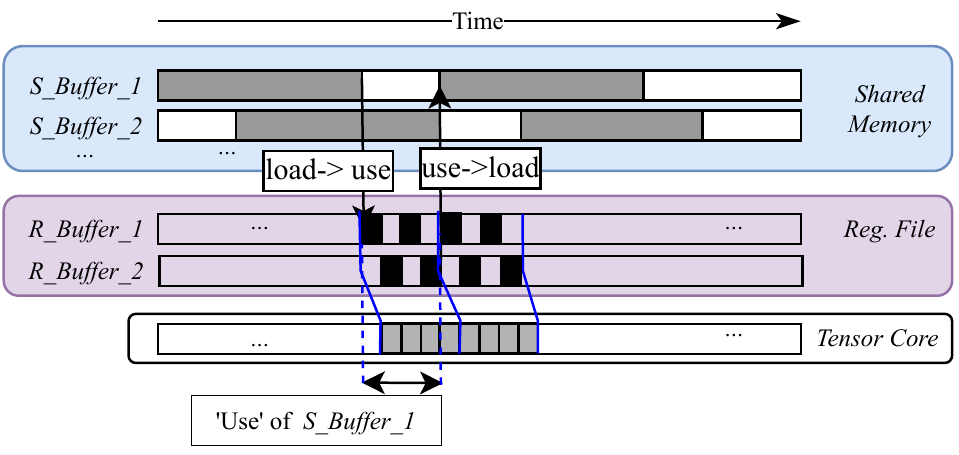}
    \caption{Multi-level pipelining with inner-pipeline fusion.}
    \label{fig:pipeline_multilevel:4}
    \end{subfigure}
    \caption{
        Concept of multi-level pipelining and inner-pipeline fusion.
        (a) shows the GPU memory hierarchy, with two levels of buffers: the shared memory and the register file.
        (b) shows the execution timeline of single-level (only shared memory) pipelining.
        (c) improves over (b) by pipelining the inner loop: register loading and computing.
        (d) improves over (c) via inner-pipeline fusion, which treats the repeated inner loop as a holistic loop and pipeline it.
        {\name} supports optimizations in (d), which provides the best performance among (b)-(d).
    }
    \label{fig:pipeline_multilevel}
\end{figure}

\begin{figure}[tb!]
    \centering
    \includegraphics[width=\linewidth]{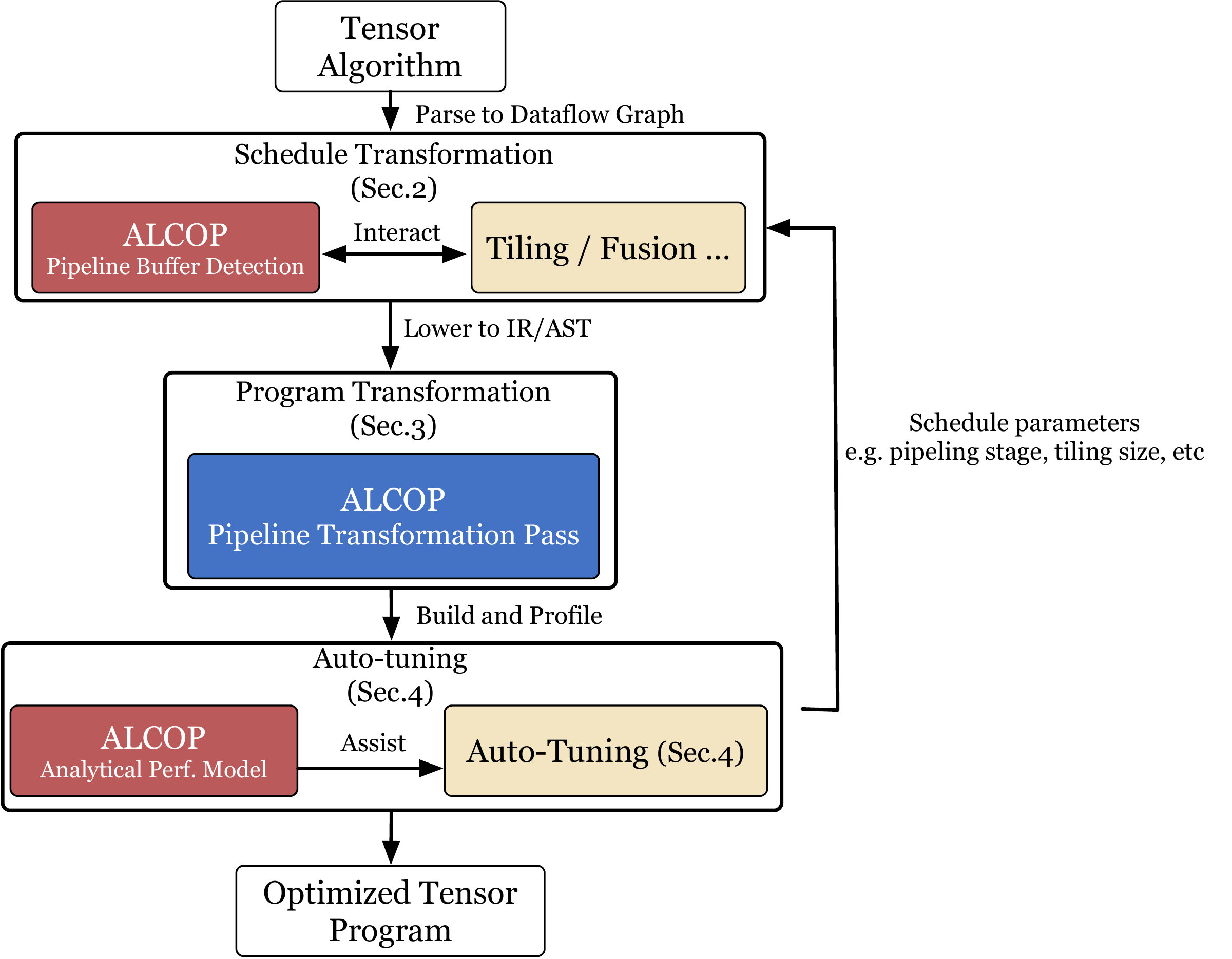}
    \caption{The overview of ALCOP.}
    \label{fig:workflow}
    \vspace{-10pt}
\end{figure}

Deep learning (DL) has achieved great success in a variety of application fields, spanning computer vision, natural language processing, and recommendation systems~\cite{he2016deep,devlin2018bert,naumov2019deep}. The widespread use of GPUs~\cite{nvidiav100,nvidiaa100} to accelerate DNNs makes an indispensable contribution in this AI era. 

High-performance tensor programs on GPUs require complex optimization efforts. When Tensor Core was introduced to GPUs to accelerate deep learning, harnessing the power of Tensor Cores became the center of GPU software optimization, motivating the development of a number of libraries and compilers~\cite{cutlass,yan2020demystifying,dakkak2019accelerating,feng2021egemm,chen2018tvm,katel2022mlir}. Because Tensor Core throughput continued to increase but memory bandwidth lagged, research on tiling and fusion to improve data re-use surged~\cite{niu2021dnnfusion,zhao2022apollo,asplos22-zhenzheng,zheng2020fusionstitching}. However, aggressively large tiling limits the number of tiles and hinders inter-tile parallelism, a crucial GPU mechanism to achieve high utilization. Restoring the parallelism lost due to aggressive tiling becomes an important task. \textbf{Pipelining} -- the overlap of data loading and computing -- is an ideal mechanism for unleashing intra-tile parallelism. Figure~\ref{fig:motivation} depicts the concept of pipelining and its performance advantages. As the difficulty of capitalizing on the ever-growing parallelism in current and future GPUs increases, the study of pipelining becomes essential.

Despite the necessity of pipelining optimization, existing approaches are either limited in their design space coverage or their degree of automation. Prior work~\cite{katel2022mlir} has studied double-buffering, a common but far less complicated case of pipelining. However, double-buffering is only a two-stage, one-level instance of the entire multi-stage, multi-level pipelining design space shown in Figure~\ref{fig:pipeline_multistage} and Figure~\ref{fig:pipeline_multilevel}, and simplifying pipelining to double-buffering hinders a major performance gain (which will be evaluated in Sec.~\ref{sec:eval:single-op}). Although deep learning systems can access comprehensive pipelining optimization from hand-written libraries~\cite{cublas} or compiler-integrated libraries~\cite{xing2022bolt}, due to their fundamental difference from the tensor program generation workflow of DL compilers, they are inextensible to new operators and indecomposable with prior compiler passes such as auto-fusion and auto-tiling. 

Automatic pipelining presents three distinct challenges: workload complexity (diverse DL operators), hardware complexity (multi-level memory hierarchy), and design space complexity (coherent performance tuning factors). 

Our key insight is that, instead of solving everything in a monolithic compiler pass, we should exploit the progressive lowering structure of DL compilers and the information exposed at each level. Specifically, we address the aforementioned three challenges through three decoupled and collaborative compilation modules: pipeline buffer detection, pipeline program transformation, and analytical-model guided design space search. Pipeline buffer detection addresses the workload complexity because it occurs during the scheduling phase when the entire dataflow is visible. The second module addresses the hardware complexity. During the program transformation stage, the intricate for-loop structure and data movement are revealed and modified. This module utilizes the safety check of the preceding module to execute the robust transformation. The third module addresses the design space complexity. It happens at the auto-tuning stage, where pipelining and other techniques are co-optimized. This module makes use of the preceding parameterized module. We further design an analytical hardware model to expedite the design space search. 

In this paper, we propose ALCOP\footnote{ALCOP is a copper-based alloy with significantly-enhanced endurance and strength. We envision our  ALCOP will significantly enhance the power of DL compilers for modern AI-GPUs.} \footnote{The source code is released at \url{https://github.com/hgyhungry/alcop-artifact}.}(\underline{A}utomatic \underline{L}oad-\underline{CO}mpute \underline{P}ipelining), the first DL compiler solution (auto-scheduler, program transformation, auto-tuner) for 
automated multi-stage, multi-level pipelining; its architecture is shown in Figure~\ref{fig:workflow}. Additional contributions include: 
\begin{enumerate}
\item We design methods to examine each buffer for applying pipelining, including the ordering of pipelining and other schedule transformations to avoid mutual interference. (Sec.~\ref{sec:schedule})
\item We design a program transformation pass that handles index manipulation, synchronization injection, and prologue injection, among other transformations. (Sec.~\ref{sec:program_transformation})
\item We propose a pipeline-aware analytical performance model. Combining it with an existing machine-learning (ML) based tuning algorithm significantly improves the efficiency of schedule tuning. (Sec.~\ref{sec:autotune})
\end{enumerate}

Experiments show that the {\name} program transformation pass can bring on average \textbf{$1.23\times$}, and up to $1.73\times$ speed-up to individual DL operators over TVM~\cite{chen2018tvm}. 
{\name}  brings {1.02-1.18$\times$} end-to-end inference speed-up for six DL models over TVM~\cite{chen2018tvm}, and 1.01-1.64$\times$ over XLA~\cite{xla}. 
Through combining the analytical model with ML-based tuning, we can identify schedules with 99\% performance compared to the best schedule in the entire design space while reducing the number of trials by $40\times$.

%% file: 3_auto_scheduling.tex
\section{Schedule Transformation}\label{sec:schedule}
Automatic pipelining begins by identifying potential pipelining possibilities. We implement it through a schedule transformation pass in the compiler, which attaches the pipelining primitive to buffer variables in the program.

Pipelining can be applied to a \textit{load-and-use} loop in which the \textit{load} step copies data into a buffer and the \textit{use} step reads data from the buffer. Consequently, the purpose of the schedule transformation is to identify and record ``load-and-use'' structures in a program. The pass marks the buffer variables within such {load-and-use} loops as pipelined buffers. Later on, a program transformation pass described in Section~\ref{sec:program_transformation} will turn the {load-and-use} structure into its pipelined version. 

Two important questions must be addressed: First, we must determine is what {rules} we should apply to identify the buffers that can be pipelined. The second one is determining the \textit{ordering} if pipelining in relation to other schedule transformations, such as tiling, aware of their mutual effect.

\subsection{Identification of Buffers for Pipelining}\label{sec:schedule:rules}

Constraints of pipelining come from not only the algorithm, \textit{i.e.,} how the buffer is used, but also the hardware capabilities, \textit{i.e.,} what forms of memory copy can be executed asynchronously. For each buffer variable, the following three rules are evaluated to determine whether pipelining can be applied.
Firstly, we do not pipeline a buffer that is not produced by asynchronous memory copy. An asynchronous memory copy indicates that the memory copy is non-blocking, so we can initiate memory copies for future loop iterations in advance and meanwhile continue with the computation in the present iteration. Only when an explicit synchronization instruction is encountered does the program block to wait for the completion of the memory copy.
If the data in a buffer is not produced by direct memory copy but rather by some compute operation, the buffer does not meet this condition.

Secondly, we do not pipeline a buffer produced outside of a \textit{sequential} loop. The purpose of pipelining is to overlap the data-loading operation of future iterations with the computation of the current iteration. This {load-and-use} loop must be sequential and cannot be parallelized (bound to parallel threads) or unrolled. 
This condition is typically violated by stencil algorithms that use tiling to increase the reuse of the input tensor, but the buffer is only filled and used once, as opposed to being generated in a sequential loop. Consequently, the pipelining approach cannot be applied to these buffers.

The final rule is about \textit{synchronizing the pipeline:} If the hardware platform supports only scope-based synchronizations, we inspect all buffers within the same scope and refuse to pipeline them if their synchronization positions do not match. Synchronizing the pipeline requires special memory barriers that await certain loading instructions (\textit{e.g.}, instructions issued in the fourth-last iteration in a 4-stage pipeline). On NVIDIA Ampere GPUs, such memory barriers are provided for the shared memory scope. Hence,  the hardware is incapable of resolving this conflict if two buffers are both in the shared memory scope, but their barriers must be inserted at distinct positions in the program. If this conflict occurs, our schedule transformation refuses to pipeline these buffers.

\subsection{Ordering of Schedule Transformations} \label{sec:schedule:order}

Pipelining is applicable to three schedule transformations already in existence: cache-reading, tiling, and fusion. We will briefly introduce these transformations and then determine whether pipelining should be applied before or after them.

\noindent{\textbf{Cache-reading.}} It means inserting a read buffer for a tensor input. Given an algorithm and computation tensor \texttt{S2} from tensor \texttt{S1}, applying cache-reading means inserting a new tensor \texttt{S1\_buf} which is an identical copy of \texttt{S1} but with a buffer scope. Cache-reading should be applied before pipelining since pipelining needs to be applied to buffers generated by the former. 

\noindent{\textbf{Tiling.}} It is the process of dividing the output tensor into blocks. In conjunction with cache-reading, it can cache data within buffers to improve data reuse.
Tiling should also be performed before pipelining. The second condition for a buffer to qualify pipelining, \textit{i.e.,} whether there exists a sequential load-and-use loop,  must be inspected based on the for-loop sketch after tiling. 

\begin{figure}[tb!]
    \centering
    \includegraphics[width=0.98\linewidth]{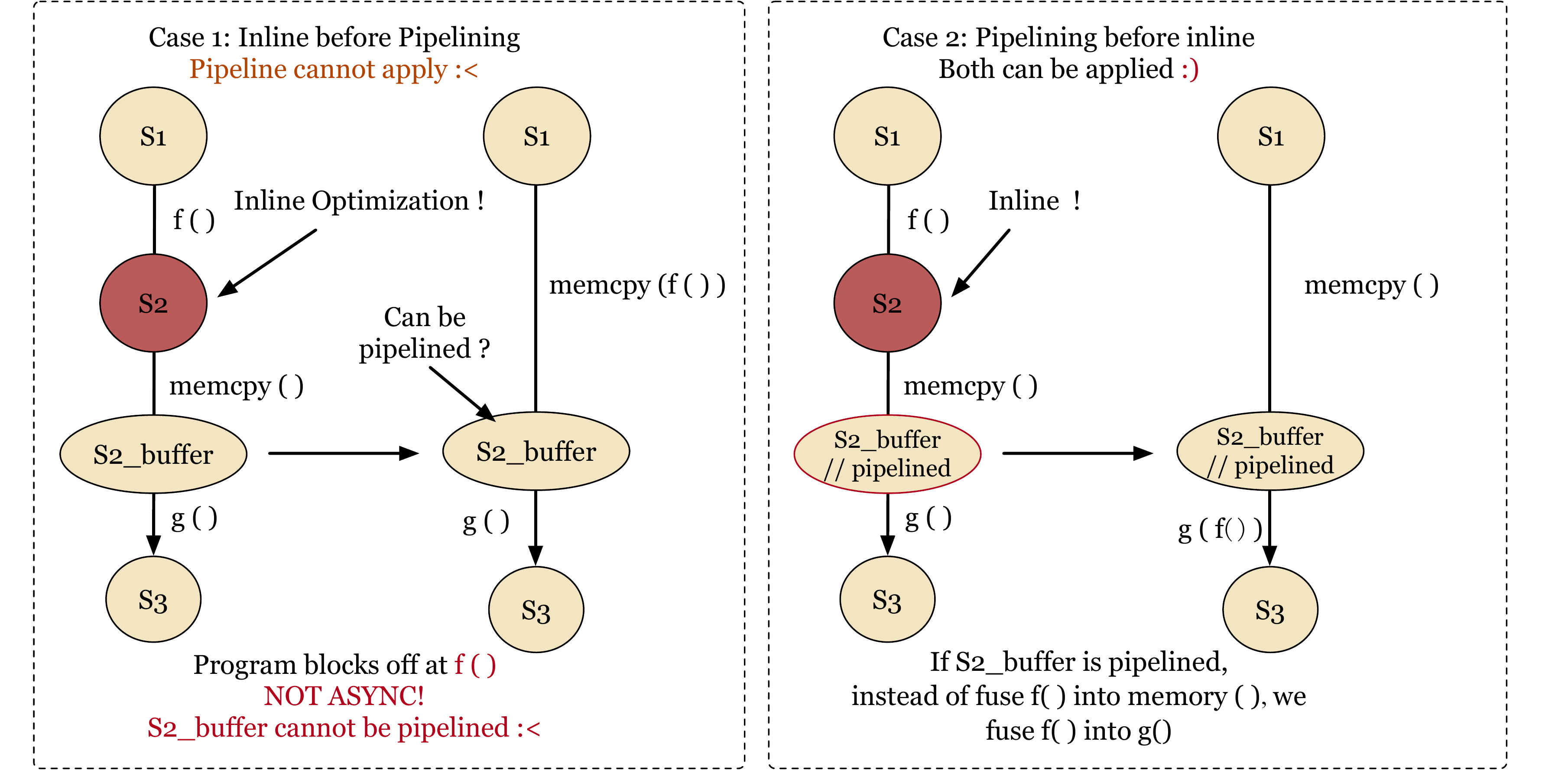}
    \caption{The effectiveness study on the optimization order of inlining and pipelining. 
    In case 1, after inlining, \texttt{S2\_buf} can no longer be pipelined because it is no longer produced by an asynchronous memory copy. In case 2, after pipelining, inlining can still be applied.}
    \label{fig:schedule_transform}
    \vspace{-10pt}
\end{figure}

\noindent{\textbf{Fusion.}} It means avoiding writing back intermediate data between two operators. Inlining, a specific type of fusion, should come after pipelining. Inlining a tensor means producing the value of the tensor precisely where it is used; this technique is often used on lightweight element-wise operations like datatype casting. Figure~\ref{fig:schedule_transform} shows an example in which originally \texttt{S2} is produced by applying element-wise function $f(\cdot)$ to \texttt{S1}, and a buffer tensor \texttt{S2\_buf} is injected after \texttt{S2}  via cache-read. Inlining \texttt{S2} is equivalent to applying $f(\cdot)$ first and then copying the data directly into \texttt{S2\_buf} without writing the data back to memory. According to our first rule outlined in the previous subsection, a pipelined memory buffer should be produced from an asynchronous memory copy. However, for \texttt{S2\_buf} here, the operation to produce it is no longer asynchronous, as the explicit $f(\cdot)$ forces the program to stall, waiting for data to be loaded. And since buffer \texttt{S2\_buf} is not produced asynchronously, it cannot be pipelined. Here in case 1, inlining impedes the opportunity of pipelining. Nevertheless, if pipelining is applied before inlining, like in case 2, the inlining of \texttt{S2} can still be applied, but in a different manner: Instead of inlining \texttt{S2} into \texttt{S2\_buf}, we cache-read \texttt{S1} and fuse the computation $f(\cdot)$ into the production of \texttt{S3}. Thus, we ensure both sides are satisfied: the buffer is produced through an asynchronous copy and can be pipelined, while computation $f(\cdot)$ is fused and we avoid explicitly generating an intermediate tensor.

\begin{figure*}[tb!]
\centering
\includegraphics[width=0.76\linewidth]{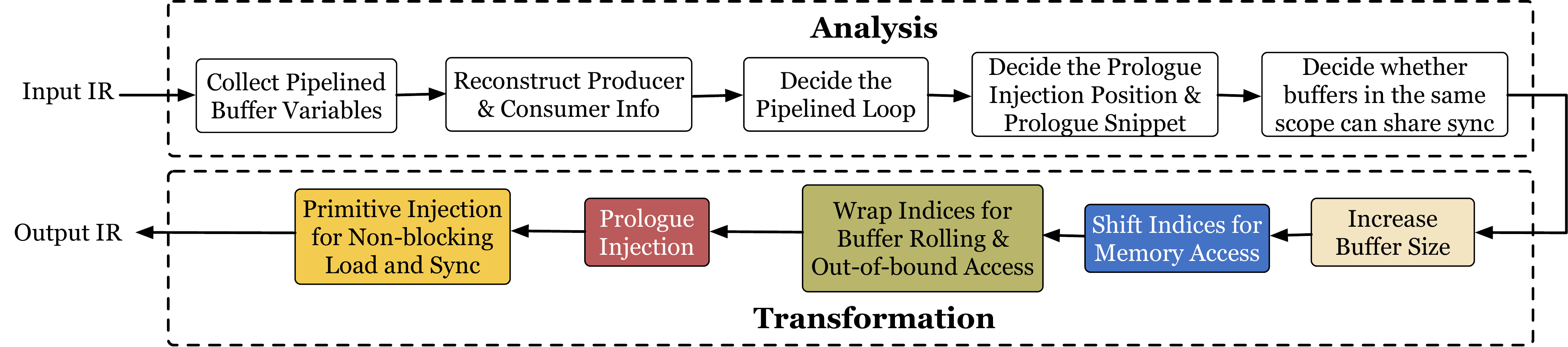}
\caption{Workflow and example input and output of the pipelining program transformation}
\label{fig:transform_workflow}
\end{figure*}

\begin{figure*}[htbp]
    \centering
    \includegraphics[width=0.36\linewidth]{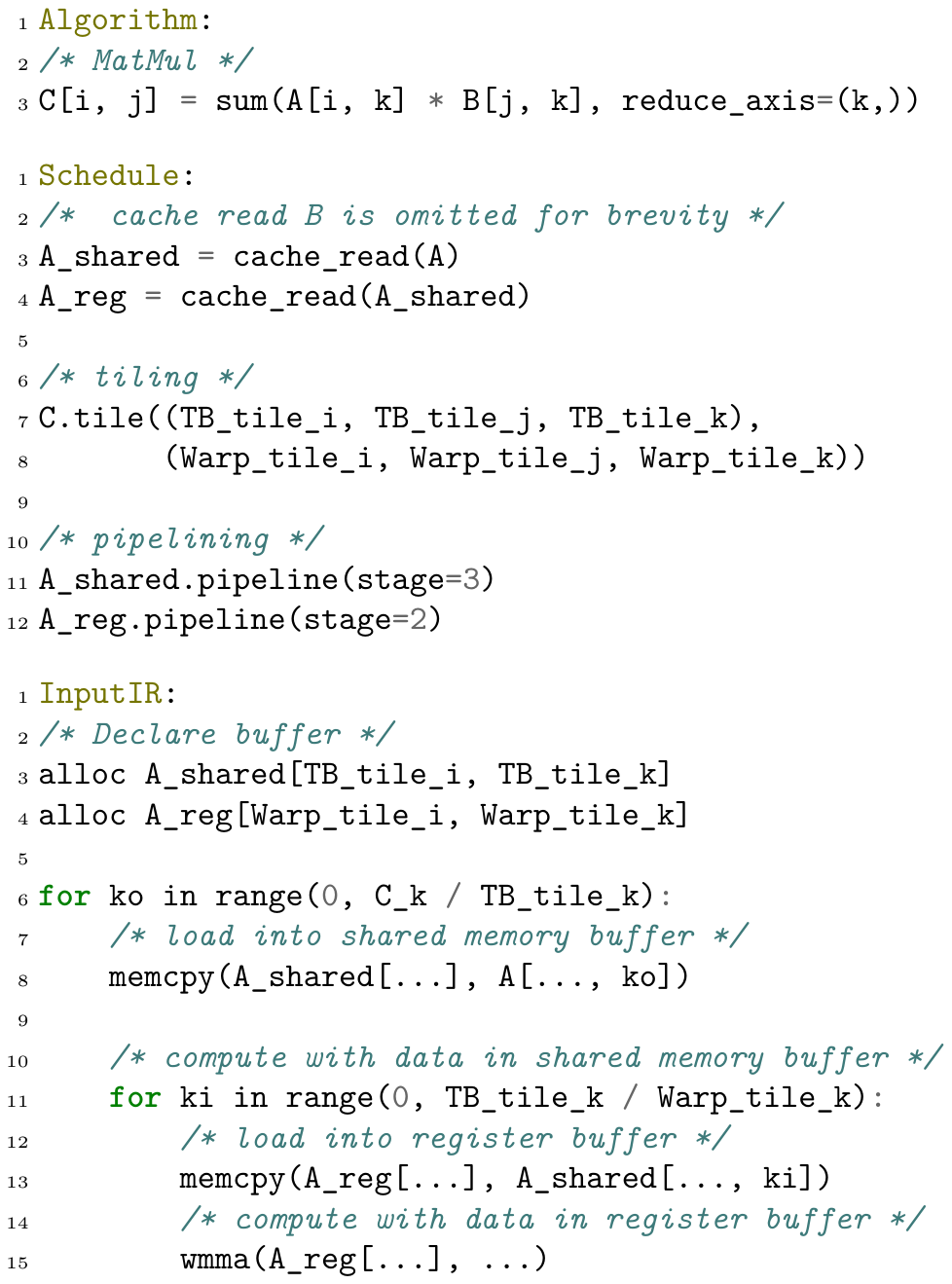}
    \hfill
    \includegraphics[width=0.58\linewidth]{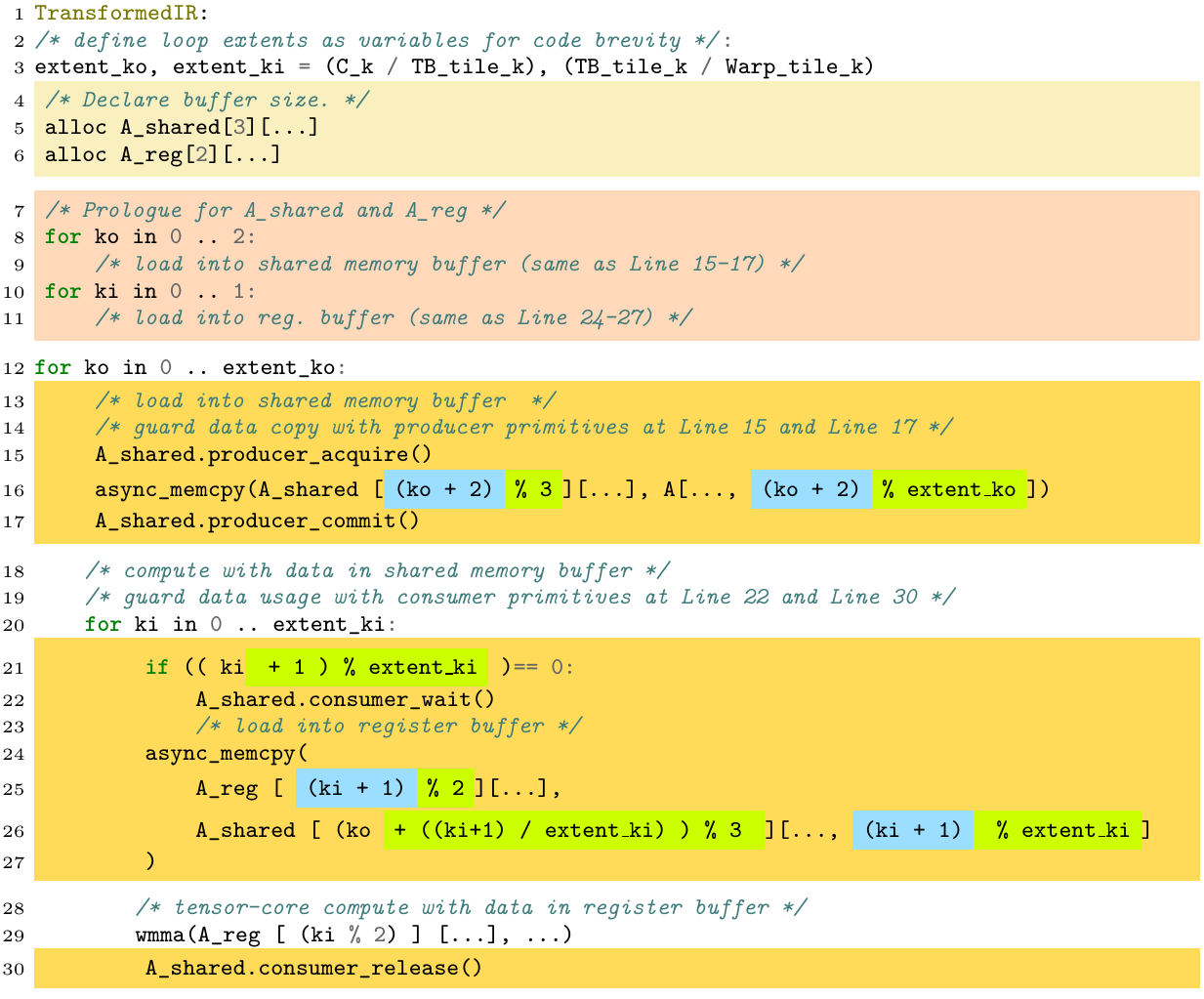}
    
    \centering
    \caption{An example to illustrate how to transform an original Tensor-IR (left) to its pipelined version (right).}
    \label{fig:code_example}
\end{figure*}

%% file: 4_v2program_transformation.tex
\section{Program Transformation} \label{sec:program_transformation}

In this section, we introduce the second component of automatic pipelining: transforming the program IR (Intermediate Representation) to implement pipelining. After the schedule transformation outlined in Section~\ref{sec:schedule}, the program is lowered to its IR form, composed of for-loops and load/store/compute operations. Figure ~\ref{fig:code_example} gives a sample input and transformed IR of the pipelining pass.
Figure~\ref{fig:transform_workflow} also depicts the transformation steps.

\subsection{Analysis}
\noindent{\textbf{The First Step.}} Given a program IR, pipelining begins with the collection of pipelining hints inserted by the schedule transformation, including the buffer to be pipelined and the number of stages for each buffer. 

\noindent{\textbf{The Second Step.}} Given a set of buffers we want to apply pipelining, the second analysis task is to reconstruct the producer tensor and consumer tensor(s) of these buffers.
Then we can derive if there are multi-level buffers by deciding if the producer of a pipeline buffer is also a pipelined buffer. 
Since pipelined buffers are always produced via asynchronous memory copy, to determine the producer tensor, it suffices to retrieve which tensor it copies from. The decision of consumers happens when IR traversal encounters a load operation from this buffer.

\noindent{\textbf{The Third Step.}} This step is to determine the sequential load-and-use loop for each pipelined buffer. This identifies the iteration variable to be pipelined and is required by all the index shifting operations in the transformation steps. The sequential loop can be determined as follows: starting from the instruction that copies data into the buffer, traversing all the for-loops from inside to outside, and finding the first sequential loop whose iteration variable is not used to index inside this buffer. This means the buffer is reused for each iteration of this loop, which is the loop we want to pipeline. Take Figure~\ref{fig:code_example} as an example, the pipelined loop for \texttt{A\_shared} is with iteration variable \texttt{ko}, and the pipelined loop for \texttt{A\_reg} is with variable \texttt{ki}. 

\noindent{\textbf{The Fourth Step.}} We should document the pieces of code that \textit{loads} and  \textit{uses} this buffer. This information is required for the injection of synchronization primitives and prologues. 
In the \textit{Input IR} in Figure~\ref{fig:code_example}, the ``loading'' part for \texttt{A\_shared} is Line \texttt{8}, and the ``using'' part is the \texttt{ki} loop and everything inside. The loading part for \texttt{A\_reg} is Line \texttt{13}, and the using part is Line \texttt{15}. 

\noindent{\textbf{The Fifth Step:}} We also need to decide where to inject prologues. Since we transform the program to issue memory copy for future iterations while doing computation for the current iteration, we need to move the first few stages of memory copy ahead of the start of the main load-and-use loop. This pre-posed loading code block is a prologue. Typically, prologues can be injected simply before the pipelined loop. However, when a multi-level pipeline appears, the prologues of inner pipelines must be injected into the sequential loop of the outer-most pipeline, in order to build a holistic pipeline as opposed to a recursive one as shown in Figure~\ref{fig:pipeline_multilevel:3}. 

\subsection{Transformations}

Five steps are required to transform a \textit{load-and-use} loop into a pipelined loop. The \textit{Transformed IR} in Figure~\ref{fig:code_example} shows the transformed version of the \textit{Input IR} in the same figure.

\noindent{\textbf{The First Step.}} This step increases the size of the memory buffer by the number of pipeline stages. Relevant transformed code is highlighted in light yellow. 

\noindent{\textbf{The Second Step.}} This step shifts the indices used in memory access. The eelevant code is highlighted in blue. In each \textit{load-and-use} iteration, we issue asynchronous memory copy for \textit{future} iterations rather than the present iteration. Therefore, we need to increase the pipelining loop variables in the memory access indices. If it is a 3-stage pipeline, for instance, we should load data 2 iterations ahead.

\noindent{\textbf{The Third Step.}} This step handles indices for buffer rolling (circular access) and out-of-bound wrapping.
Relevant code is highlighted in green. There are two cases that we need to wrap indices: first, when we use the pipelining variable to index a chunk of the buffer, we should use the modulo of pipeline iteration variable divided by pipeline stages. 
Secondly, since we increase the pipelining variable, it is possible that we index out of the bound of its producer tensor. We must take the modulo of the pipelining variable divided by its own extent to avoid indexing out-of-bound.
A complicated case is in a multi-level pipeline when the overflow of the inner pipeline causes the increase of the outer pipeline variable. Line \texttt{26} in the transformed IR handles this case.

\noindent{\textbf{The Fourth Step.}} This step injects prologue primitives. The contents of prologues are the memory copy of the first \texttt{n\_stage -1} chunks of data, where \texttt{n\_stage} is the number of the pipeline stage. We inject prologue at the positions we record in the preceding analysis pass.

\noindent{\textbf{The Fifth Step.}} The final step injects synchronization primitives. The pipeline is guarded by four primitives: \texttt{producer\_acquire}, \texttt{producer\_commit}, \texttt{consumer\_wait}, and \texttt{consumer\_release}. \texttt{producer\_commit} commits a batch of asynchronous loading operations. \texttt{consumer\_wait} blocks until a previous batch of loading is completed.  When the pipeline is full, \texttt{producer\_acquire} blocks until \texttt{consumer\_release}{\footnote{{\url{https://docs.nvidia.com/cuda/cuda-c-programming-guide/index.html\#with-memcpy_async-pipeline-pattern-multi}}}}. The pairs of producer/consumer primitives are put around the loading/using part of the buffer, respectively, as shown in Line \texttt{15, 17, 22,} and \texttt{30} of the transformed IR.

%% file: 5_v2perfmodel.tex
\section{\hpcarevise{Static Analysis Guided Tuning}}\label{sec:autotune}
\begin{figure}
    \centering
    \includegraphics[width=0.9\linewidth]{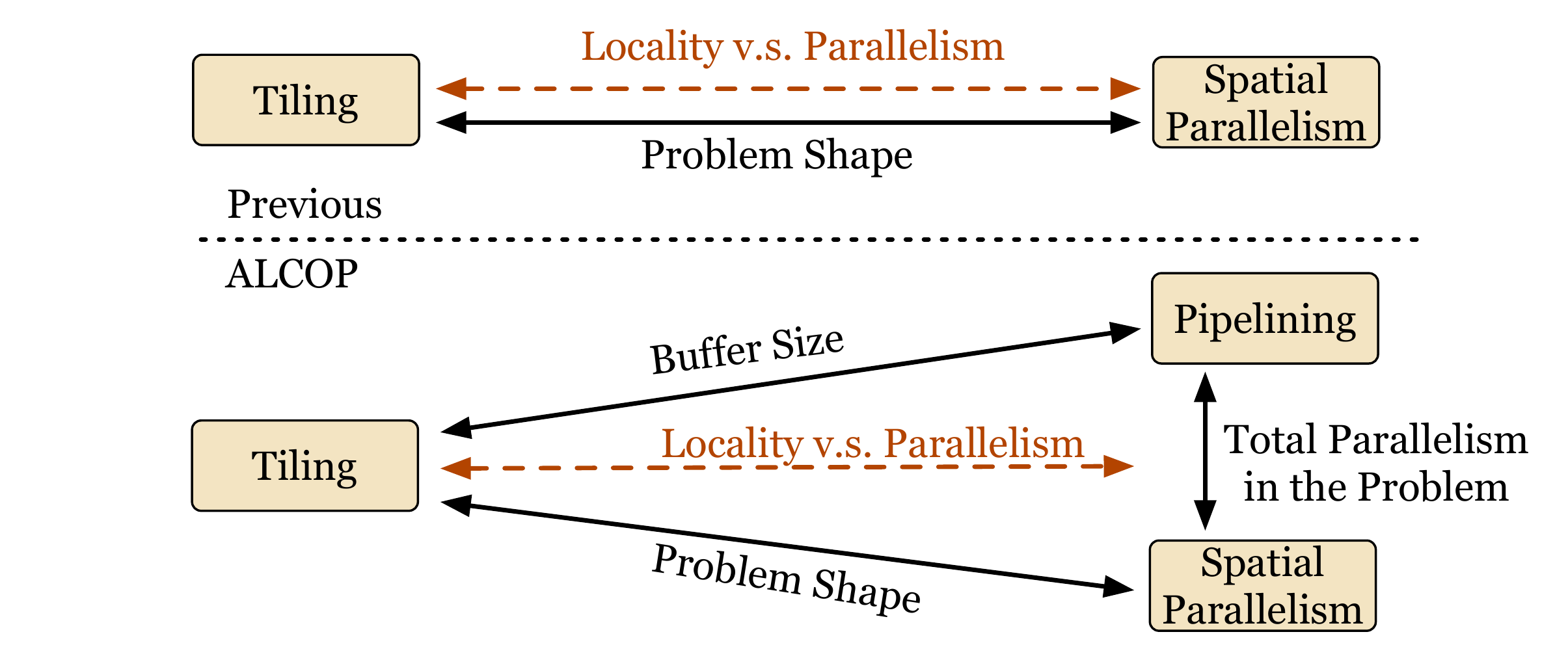}
    \caption{\hpcarevise{A high-level view of the performance model. Compared to prior work~\cite{lym2019delta}, our model takes into account the constraints and trade-offs among pipelining, tiling and spatial parallelism.}}
    \label{fig:perf_model_high_level}
    \vspace{-10pt}
\end{figure}

This section introduces how we combine a static analytical performance model with existing machine-learning (ML) based auto-tuning~\cite{chen2018learning} to choose schedule parameters. The key component is a novel performance model aware of pipelining and its interaction with other optimizations, as illustrated in Figure~\ref{fig:perf_model_high_level}. 

\subsection{Top-Level Model}

\begin{table*}[t]
\caption{Analytical Performance Model}
\resizebox{\linewidth}{!}{
\begin{tabular}{|l|l|}
\hline
\textbf{Category}         & \textbf{Model}                                                                                                                                                                                                                                                                                                                                                                                                                       \\ \hline \hline
Kernel Latency Model           & $T_{kernel} = T_{threadblk} \times N_{threadblk\_batch}$                                                                                                                                                                                                                                                                                                                                                                                    \\ \hline
Pipeline Latency Model      & \begin{tabular}[c]{@{}l@{}}Input: $T_{load}, T_{use}, N_{loop}, N_{pipe}, N_{mplx}$\\ Output: $T_{load\_use\_loop}$\\ If $T_{load} \leq (N_{pipe} \times N_{mplx} -1) \times T_{use}$: $T_{load\_use\_loop} = T_{use} \times N_{loop}$\\ Else: $T_{load\_use\_loop} = (T_{load} + T_{use}) \times N_{loop} \div N_{pipe}$\end{tabular}                                \\ \hline
Threadblock Latency Model & \begin{tabular}[c]{@{}l@{}}$T_{threadblk} = T_{init} + T_{main\_loop} + T_{epilogue}$\\ $T_{init} = T_{smem\_load} + T_{reg\_load}$\\ $T_{main\_loop} = \texttt{PipelineLatencyModel}(T_{smem\_load}, T_{smem\_use}, N_{smem\_loop}, N_{smem\_pipe\_stage}, N_{threadblk\_per\_SM})$\\ $T_{smem\_use} = \texttt{PipelineLatencyModel}(T_{reg\_load}, T_{compute}, N_{reg\_loop}, N_{reg\_pipe\_stage}, N_{warp\_per\_threadblk})$\end{tabular} \\ \hline
Computation Latency Model & $T_{compute} = \cfrac{FLOPs_{one\_reg\_loop}}{Throughput_{SM } \times \texttt{Util}(N_{warp\_per\_threadblk}, N_{threadblk\_per\_SM})}$                                                                                                                                                                                                                                                                                                        \\ \hline
Memory Latency Model      & \begin{tabular}[c]{@{}l@{}}$T_{snen\_load} = MAX(T_{LLC\_load}, T_{DRAM\_load}) $\\ $T_{LLC\_load} = LAT_{LLC\_read} + \cfrac{Bytes_{one\_smem\_loop} \times N_{threadblk\_per\_threadblk\_batch}}{BW_{LLC}}$\\ $T_{DRAM\_load} = LAT_{DRAM\_read} + \cfrac{Bytes_{threadblk\_batch\_workset} }{BW_{DRAM}}$\end{tabular}                                                                                                                             \\ \hline
Epilogue Model            & $T_{epilogue} = LAT_{DRAM\_write} + \cfrac{Bytes_{output\_tile} \times {N_{threadblk\_per\_threadblk\_batch}}}{BW_{DRAM\_write}}$                                                                                                                                                                                                                                                                                                            \\ \hline
\end{tabular}
}
\label{tb:perfmodel}
\end{table*}

Our analytical model is shown in Table~\ref{tb:perfmodel}. At the top level, the threadblocks are grouped into threadblock-batches ($threadblk\_batch$), and one threadblock-batch occupies all Streaming Multiprocessors (SMs) at a time. Since all threadblocks execute the same program, the latency of a kernel equals the threadblock latency multiplied by the number of batches.
The number of threadblock-batches in a kernel depends on the GPU scheduling policy, which we learn through performance profiling. 
The maximum number of threadblocks per SM is limited by the size of shared memory and register files that each SM can provide, as well as the request of threadblock. Our simulated GPU scheduling policy considers all these factors to decide $N_{threadblk\_batch}$.

At the threadblock level, we estimate its final performance by summing the latencies of three phases: (1) the initial phase $T_{init}$, in which the first chunk of data is requested and the pipeline waits for it to arrive; (2) the main loop $T_{main\_loop}$, in which the {load-and-use} pipeline advances at a steady rate; (3) the epilogue phase $T_{epilogue}$, in which the final results are written back into the global memory. $T_{epilogue}$ is determined using the Epilogue Model equation proposed in DELTA~\cite{lym2019delta}. 

Let us consider $T_{main\_loop}$. It illustrates {load-and-use} loop at the shared memory level, which comprises copying data from the device memory to the shared memory, reading the data into the register, and doing computations with tensor cores. We employ a Pipeline Latency Model, which is described in the next subsection, to calculate the latency of the loop. This model considers the pipelining and multiplexing factors, $N_{pipe}, N_{mplx}$, which means the number of stages the pipeline has, and the number of parallel workers that can be multiplexed to hide the memory copy latency. At the shared memory level, these two parameters equal to the number of stages at the outer {load-and-use} loop, $N_{smem\_pipe\_stage}$, and the number of parallel threadblocks in an SM, $N_{threadblk\_per\_SM}$. 

Calculation of $T_{main\_loop}$ still needs the latency of the use phase in this loop. However, the use phase is another pipeline that loads data into the register files and performs computations with tensor cores. We can calculate the latency of the use phase by estimating the stable state latency of the inner pipeline through inner-pipeline fusion. For this inner load-and-use loop, the use latency refers to the latency of performing arithmetic operations inside one loop on tensor cores. The pipeline and multiplex factors are determined by the number of stages of this inner load-and-use loop and the number of parallel warps in a threadblock.

\subsection{Obtaining Detailed Latencies}
\begin{figure}[tb!]
    \centering
    \begin{subfigure}{\linewidth}
    \centering
    \includegraphics[width=0.85\linewidth]{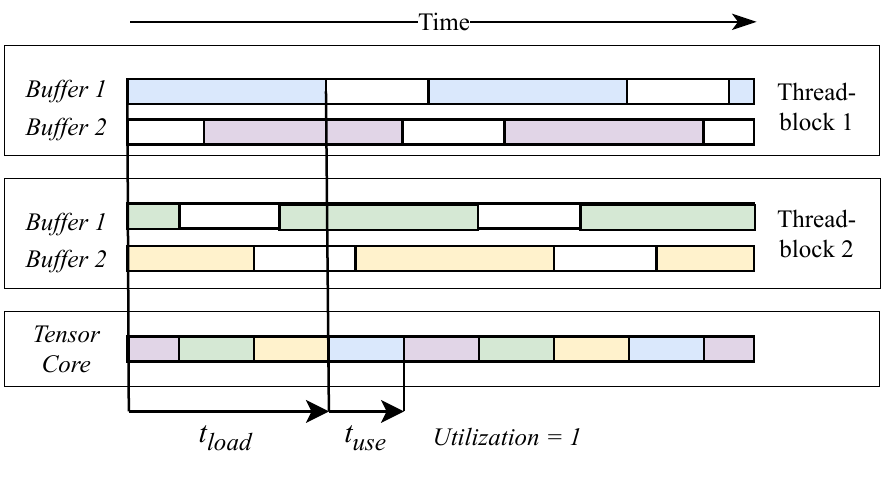}
    \caption{Case 1: $t_{use}\cdot N_{mplx} \cdot N_{pipe} \geq (t_{load} + t_{use})$
    }
    \label{fig:pipeline_latency_1}
    \end{subfigure}
    \begin{subfigure}{\linewidth}
    \centering
    \includegraphics[width=0.85\linewidth]{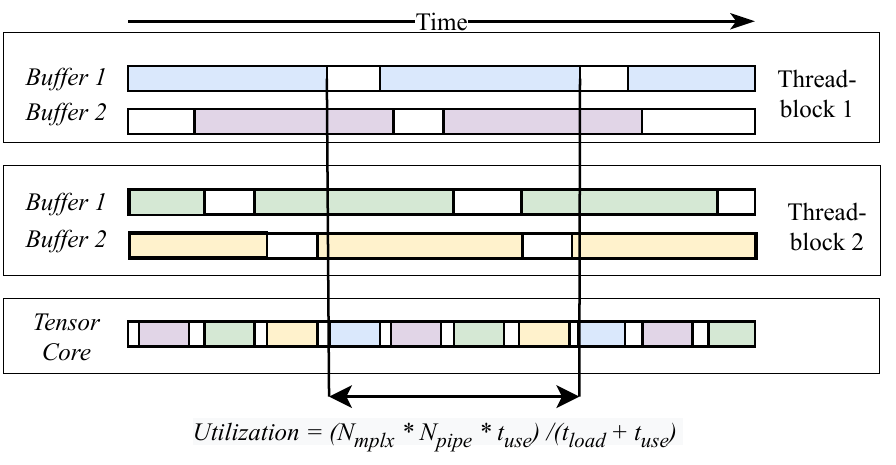}
    \caption{Case 2: $t_{use} \cdot N_{mplx} \cdot N_{pipe}  < (t_{load} + t_{use})$
    }
    \end{subfigure}
    \caption{Explanation of the pipeline latency model. A \textit{load} can be  overlapped by \textit{computing} in other threadblocks, or in other stages of the same threadblock. }
    \vspace{-10pt}
    \label{fig:pipeline_latency}
\end{figure}

\begin{figure*}[tb!]
    \centering
    \includegraphics[width=0.95\linewidth]{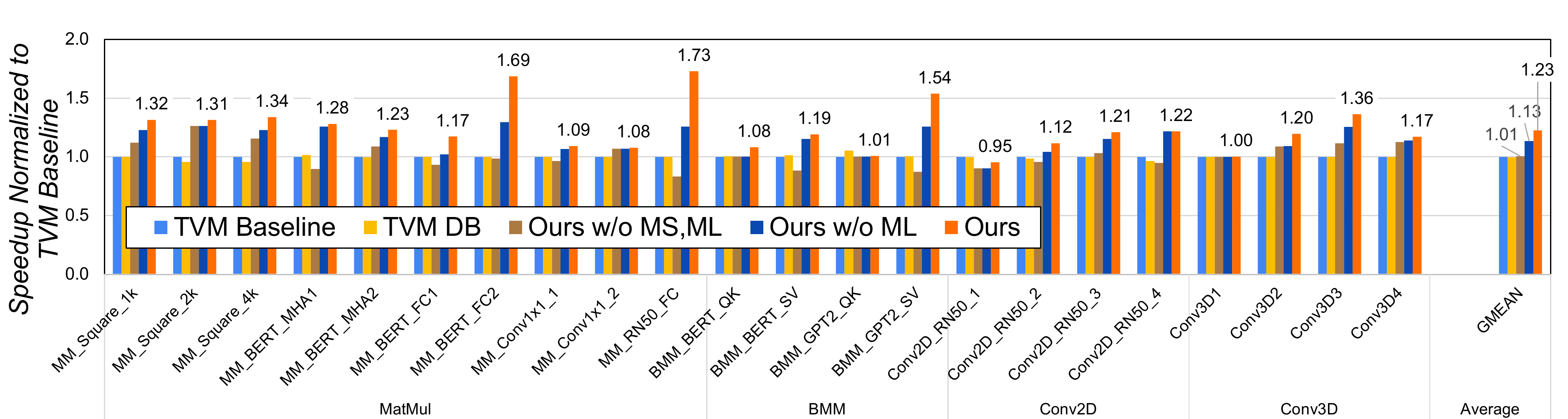}
    \caption{Single operator performance normalized to TVM on A100.}
    \label{fig:single_op_a100}
    \vspace{-10pt}
\end{figure*}

\noindent{\textbf{Pipeline Latency Model.}}
Now we address the core issue of estimating the latency of a load-and-use loop in its stable state. Intuitively, the prediction should differ depending on whether the bottleneck is loading or using.
Line 3 in the Pipeline Latency Model in Table~\ref{tb:perfmodel} is the criterion for determining the bottleneck. Figure~\ref{fig:pipeline_latency} illustrates the two scenarios in which computation or loading is the bottleneck. The intuition is that, during the loading of one data chunk, the computation units can be used to compute other chunks of data in this pipeline ($N_{pipe}$), or used for other parallel workers ($N_{mplx}$). If the latency of data loading exceeds the latency of all computations that can overlap with it, the loading becomes the bottleneck, making the loop latency equal to the latency of one load-and-use iteration, divided by the number of overlapping streams, i.e., ${(T_{load}+T_{use})}/{N_{pipe}}$. 

\noindent{\textbf{Computation and Memory Latency Model.}}
To obtain the computation latency, we can simply divide the number of float-point operations performed inside a loop by the tensor core throughput in an SM. 
When determining the latency of memory copies, four parameters must be considered: the amount of data transferred, the available bandwidth, the number of parallel workers (threadblocks or warps) to share with the bandwidth, and a constant round-trip latency $LAT$ . Note that GPU LLC is shared by all SMs. Hence the DRAM traffic cannot be computed by the sum of data loaded by all threadblocks because the data may hit in LLC. We model DRAM traffic by deciding the working set of a threadblock-batch.

\subsection{{Model-Guided Auto-Tuning}}\label{sec:autotune:model-guided-tuning}

Now, we will discuss how to use the analytical performance model for scheduled tuning. The workflow of auto-tuning is composed of a \textit{cost model} to predict performance from schedule, and a \textit{sampling method} to propose new \textit{trials}. Unlike the analytical model we developed, TVM uses not an analytical model but a machine learning (ML)-based cost model that only learns from the profiled performance results.
Analytical model and ML-based tuning offer complementary benefits: analytical model does not require the complexity of compiling and running sampled schedules but cannot be very accurate because it is difficult to capture hardware factors such as memory system thoroughly. ML-based tuning learns the cost model from measured performances that incorporate these complex factors, but it requires a large amount of sampled data, leading to a lengthy tuning process. 

Finally, we leverage the analytical performance model's prediction to pre-train the ML-based model, allowing the ML model to acquire previous knowledge while still utilizing profiled data. 
Table~\ref{tb:search_methods} compares our method (Model-Assisted XGB) with other available auto-tuning approaches. 

\begin{table}[tb!]
\caption{{Comparison of compiler search methods.}}
\resizebox{1.06\linewidth}{!}{
\begin{tabular}{|c|c|c|c|c|}
\hline
                   & \textbf{\begin{tabular}[c]{@{}c@{}}Grid\\ Search\end{tabular}} & \textbf{XGB}                                                  & \textbf{\begin{tabular}[c]{@{}c@{}}Anal.\\ Only\end{tabular}} & \textbf{\begin{tabular}[c]{@{}c@{}}Anal.\\ + XGB (ours)\end{tabular}} \\ \hline \hline
Cost Model         & \multirow{3}{*}{N.A.}                                          & ML                                                            & Analytical                                                    & ML                                                             \\ \cline{1-1} \cline{3-5} 
Prior Knowledge?   &                                                                & No                                                            & Yes                                                           & Yes                                                            \\ \cline{1-1} \cline{3-5} 
Update Cost Model? &                                                                & Yes                                                           & No                                                            & Yes                                                            \\ \hline
Sampling           & Enumerate                                                      & \begin{tabular}[c]{@{}c@{}}Simulated\\ Annealing\end{tabular} & \begin{tabular}[c]{@{}c@{}}Cost-Model\\ Ranking\end{tabular}  & \begin{tabular}[c]{@{}c@{}}Simulated\\ Annealing\end{tabular}  \\ \hline
\end{tabular}
}
\label{tb:search_methods}
\vspace{-10pt}
\end{table}

%% file: 6_evaluation.tex
\section{Evaluation} \label{sec:evaluation}



\subsection{Single Operator Performance}\label{sec:eval:single-op}
This part evaluates pipelining speedup on single operators. Our benchmarks extracted from real DNN workloads contain four operators with a variety of shapes. All operators use half-precision and run on Tensor Cores.
We run all experiments on NVIDIA Ampere GPU, as prior generations lack the asynchronous memory-copy hardware feature. Our evaluation platform is NVIDIA A100-SMX4 with 40GB device memory. The software we use is CUDA v11.4.

We implement our pipelining framework based on TVM~\cite{chen2018tvm} v0.8 and compare it against the vanilla TVM. We augment both ALCOP and baselines with shared memory swizzling to avoid bank conflict limitation. 
We also manually insert double-buffering primitives into TVM and use it as the second baseline (TVM DB). We also compare against two downgraded versions of our compiler for ablation study: ALCOP without multi-level (ML), meaning just pipelining in shared memory level, and ALCOP without ML and multi-stage (MS), meaning only allowing two-stage pipelining. Here we exhaustively search the schedule space and give the best schedule for ours and all baselines. 

Figure~\ref{fig:single_op_a100} shows the performance of different compilers normalized to TVM. Our compiler produces operators that are 0.95-1.73$\times$, on average 1.23$\times$, faster than TVM. 
Pipelining is especially effective for operators with small output shapes but long reduction axis. Take matrix-multiplication (MatMul) as an example, MM\_RN50\_FC, the operator that gives the largest speedup, has an output shape of $1024\times 64$, and a reduction axis of $2048$. Also, for Batched Matrix Multiplication (BMM), the operators with short reduction axis (e.g., BMM\_BERT\_QK) show much smaller speedup than those with long reduction axis (e.g., BMM\_BERT\_SV).

\noindent{\hpcarevise{\textbf{Insights about when pipelining works well.}} }
\hpcarevise{Problems with small output shapes (\textit{e.g.,} MM\_BERT\_FC2, MM\_RN50\_FC) have limited spatial parallelism, so they benefit more from pipelining since pipelining uncovers extra parallelism. For problems with large output shapes (\textit{e.g.,} MM\_Conv1x1\_1), or with small reduction dimensions (\textit{e.g.,} BMM\_GPT2\_QK), pipelining provides limited benefit since the former already have abundant parallelism and the latter cannot amortize the latency of initial loading stages in the pipelining schedule.}

\noindent{\textbf{Ablation study.}} 
Multi-level and multi-stage pipelining are both critical to final speedup. As shown in Figure~\ref{fig:single_op_a100}, TVM DB  does not bring obvious speedup over TVM. Without multi-level pipelining, ALCOP can only provide an average 1.13$\times$ speedup. Without multi-level and multi-stage pipelining, ALCOP can only give $1.01\times$ speedup over TVM. 

\subsection{End-to-End Performance}\label{sec:eval:e2e}


To evaluate end-to-end model acceleration, we compare against two baselines: TVM~\cite{chen2018tvm} and XLA~\cite{xla} (TF v2.9.1). XLA is a compiler integrated into the Tensorflow framework to optimize models in an end-to-end fashion. We evaluate six popular deep learning models. 
BERT, BERT-Large~\cite{devlin2018bert} and GPT-2~\cite{radford2019language} are popular models in Natural Language Processing (NLP). ResNet-18, ResNet-50~\cite{he2016deep} and VGG-16~\cite{simonyan2014very} are three convolution neural networks widely used in vision tasks. 
Pipelining can be applied to MatMuls, BMMs and Conv2Ds, which are the most computation intensive operators and consumes a great proportion of the inference latency in these models. 

Table~\ref{tb:end2end} shows the end-to-end speedup in real models. 
We achieve 1.02-1.18$\times$ end-to-end speedup over TVM and 1.01-1.64$\times$ speedup over XLA. 

\begin{table}[tb!]
    \centering
    \caption{Model speedup from pipelining}
    \label{tb:end2end}
    \resizebox{.88\linewidth}{!}
    {
        \begin{tabular}{|c|c|c|}
            \hline
            \textbf{Model} & \textbf{Speedup over TVM} & \textbf{Speedup over XLA} \\ \hline \hline
            BERT           & 1.15                      & 1.27                      \\ \hline
            BERT-Large     & 1.18                      & 1.16                      \\ \hline
            GPT-2          & 1.15                      & 1.34                      \\ \hline
            ResNet-18      & 1.02                      & 1.64                      \\ \hline
            ResNet-50      & 1.06                      & 1.02                      \\ \hline
            VGG-16         & 1.10                      & 1.01                      \\ \hline
        \end{tabular}
    }
\end{table}

\subsection{Comparison with Libraries} \label{sec:eval:library}

\begin{figure}[tb!]
    \centering
    \includegraphics[width=\linewidth]{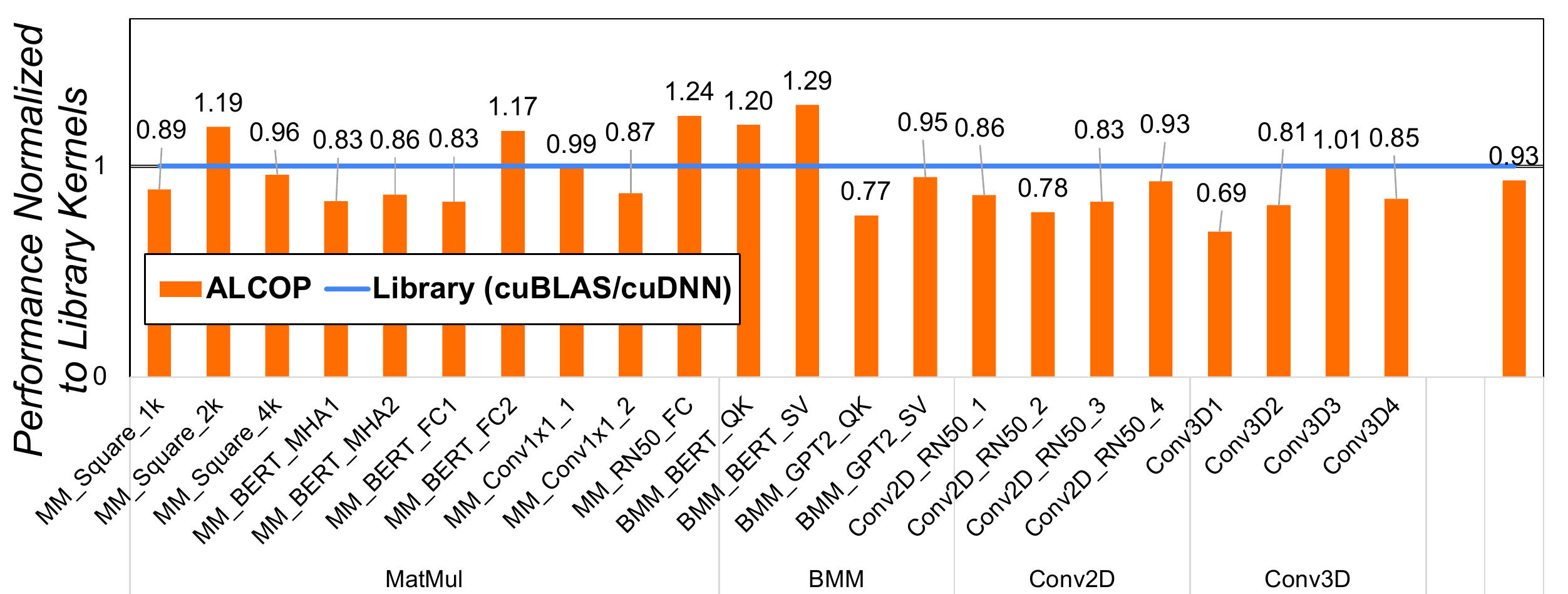}
    \caption{Single operator performance versus libraries.}
    \label{fig:single_op_library}
\end{figure}

We compare with kernels in vendor libraries (cuBLAS \cite{cublas}/cuDNN~\cite{cudnn}), which are heavily hand-optimized for the typical problem shapes we evaluate. Note that despite their high performance, libraries take huge manual efforts due to low modularity and cannot replace compilers in AI-GPU optimization.


Figure~\ref{fig:single_op_library} shows the performance of ALCOP normalized to library kernels. We can achieve on-par, on average 93\% normalized, performance compared with library kernels. For some operators like BMM\_BERT\_QK, our compiler even generates faster kernels than cuBLAS because our compiler can search the entire schedule space and find the best schedule for input operators.

\subsection{Performance Model Accuracy}\label{sec:eval:perfmodel}
\begin{figure}[tb!]
    \centering
    \includegraphics[width=\linewidth]{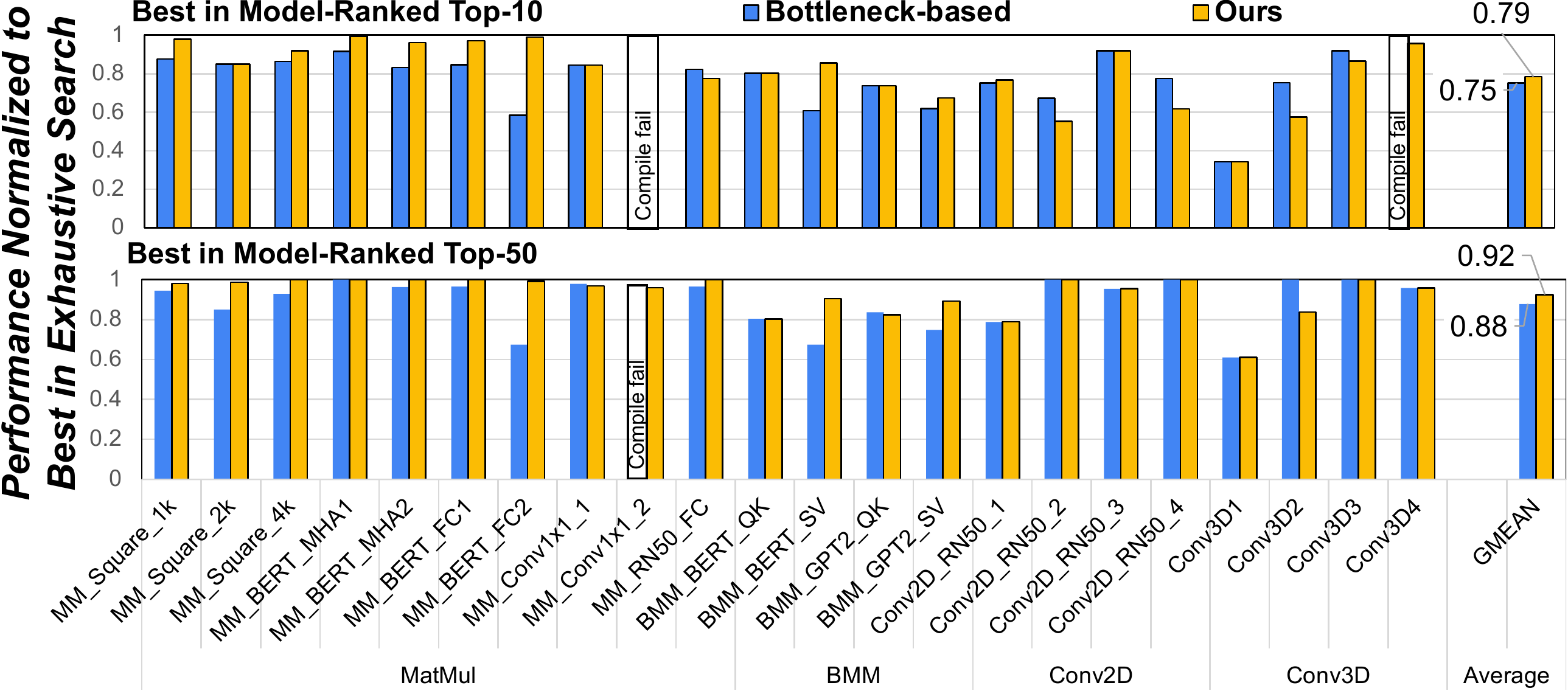}
    \caption{Best-in-top-$k$ performance of two analytical performance models. \hpcarevise{The mark 'compile fail' means the first 10 or 50 proposed schedules fail to compile into executables.}}
    \label{fig:perf_model1}
    \vspace{-10pt}
\end{figure}

\begin{figure*}[tb!]
    \centering
    \includegraphics[width=0.65\linewidth]{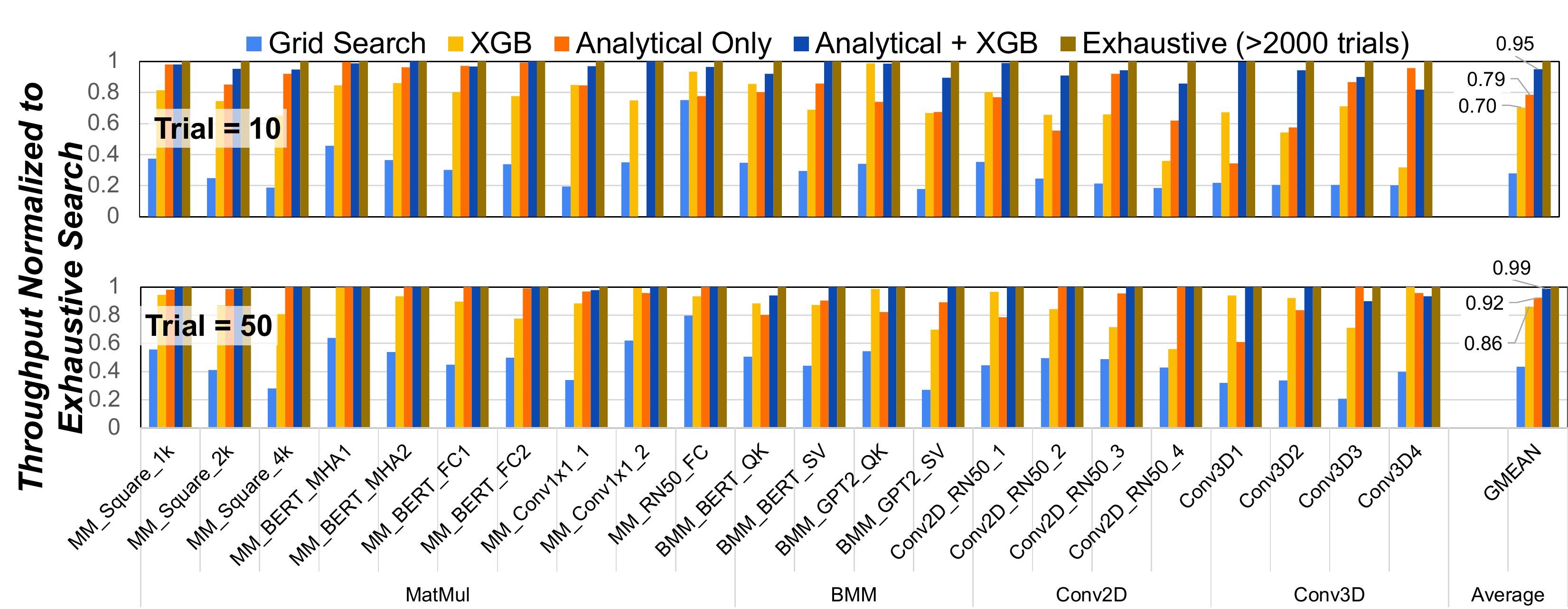}
    \caption{Search efficiency of schedule tuning methods. }
    \label{fig:perf_model2}
\end{figure*}

The metric we use to evaluate our performance model is {best performance in model-ranked top-$k$ schedules}, or \textit{best-in-top-$k$} in short. It means the best performance within the top $k$ schedules is predicted by the performance model. Compared to mean-absolute-error among the entire schedule space, best-in-top-$k$ is more meaningful to schedule tuning because tuning cares about finding efficient schedules within a limited number of trials. 

We compare against bottleneck-based analysis, a simple model that takes the maximum of computation, shared memory loading and device memory loading time, assuming full utilization of computation throughput and bandwidth. It is over-simplified in the following ways: (1) assumes an aggregated computation unit, but in GPUs the Tensor Cores are distributed in different SMs and occupancy of SMs matters. (2) agnostic to the latency hiding effect, which is what pipelining mainly benefits. 


Figure~\ref{fig:perf_model1} shows the best-in-top-$k$ results for our analytical model and bottleneck-based analysis for $k=10,k=50$. All results are normalized to exhaustive search, \textit{i.e.}, the best performance in the entire schedule space. Within the top-10 trials, our performance model achieves an average of 79\% performance compared to the best in exhaustive search, but the bottleneck-based method only achieves 75\%. Within the top-50 trials, which is a 40$\times$ saving of trials compared to exhaustive search,  our model achieves an average 92\% performance, whereas the bottleneck-based method only achieves 88\%. Our model also achieves $>$95\% performance for all matrix-multiplication (MatMul) operators.

\subsection{Analytical-Model-Guided Schedule Tuning}\label{sec:eval:autotuning}

This part evaluates our technique to combine the analytical model with machine learning (ML) based schedule-tuning. 
The metric is \textit{best-in-$k$-trials} similar as in the last part. We compare our method with the other three methods, as detailed in Table~\ref{tb:search_methods}:
(1) Grid-Search, which simply grid-search all the parameter configurations and does not learn anything from the collected performance data.
(2) XGB, which is the default method in TVM~\cite{Tavarageri2021}, and uses XGBoost~\cite{chen2016xgboost} as a cost model to fit the collected data and uses simulated annealing to propose new trials.
(3) Analytical-only, which ranks all schedules according to their \textit{predicted} performance via our analytical model
(4) Analytical$+$XGB, which first pretrains the XGB model offline with pairs of schedules and their predicted performance from the analytical model, and next follows the same workflow as XGB.


Figure~\ref{fig:perf_model2} shows the \textit{best-in-$k$-trials} of the four searching methods, normalized to the best performance in exhaustive search. At a budget of 10 trials, our Model-Assisted XGB finds schedules that reach 95\% of the best performance in exhaustive search, while sampling purely based on an analytical model or non-pretrained XGB gives 79\% and 70\% of the best possible performance accordingly. At a budget of 50 trials, which is a $>40\times$ saving of trials compared to an exhaustive search, our method reaches 99\% of the best possible performance, while Model-Ranking and XGB only obtain 92\% and 86\%, respectively. To sum up, we find that (1) analytical model helps ML: Model-Assisted XGB is better than XGB because it incorporates prior knowledge about the hardware, and (2) ML helps analytical model: Model-Assisted XGB is better than pure analytical model because it uses the actual profiled data to fine-tune the performance model. 

%% file: 7_v2relatedwork.tex
\section{Related Work}

\noindent\textbf{Pipelining.}
Pipelining, as a GPU kernel optimization, is frequently used in GPU libraries like CUTLASS~\cite{cutlass}. CUTLASS implements pipelining in matrix multiplication and convolution kernels. However, being a template-based kernel library, CUTLASS is unable to provide automatic pipelining for any tensor programs; this is only possible with our compiler-based solution. 

The term ``pipelining'' in distributed DL training~\cite{huang2019gpipe,narayanan2019pipedream,barham2022pathways,zheng2022alpa} refers to  operator-wise parallelism. In this case, different GPUs compute different stages of a model, and multiple micro-batches are computed in a pipelined fashion. Compared to our optimization at the scope of a single kernel, those model-level pipelining work use distinct techniques and focuses mainly on stage partitioning strategy. 
Pipelining is also a hardware design technique widely used in accelerator designs~\cite{liu2016cambricon,jouppi2017datacenter,sohrabizadeh2020end,liao2021ascend}, or hardware generation languages~\cite{wei2017automated,lai2019heterocl,wang2021autosa,parashar2019timeloop}. Despite sharing the same mission of improving computation and memory system utilization, the hardware-based pipelining and our compiler-based approach are very different in that hardware technique mainly benefit accelerator design, but our technique benefits program optimization for general-purpose architectures like GPU. 
\hpcarevise {Software-pipelining has been studied to exploit instruction-level parallelism~\cite{ning1993novel,govindarajan1996framework} and multithread parallelism~\cite{wei2012software}. Compared to those, ALCOP's task is more challenging because it must support multi-level pipelining and must automatically split code into a load- or compute-blocks using IR analysis, whereas, in other settings, the pipeline stages are straightforward.}

\noindent\textbf{Performance Model.}
There is a rich amount of work on analytical performance models for GPUs~\cite{hong2009analytical,volkov2016understanding,wang2020mdm,huang2014gpumech,zhang2011quantitative,baghsorkhi2010adaptive,lym2019delta}. The most relevant is DELTA~\cite{lym2019delta}, which builds a model to predict the latency of Conv2D kernels on GPUs. \hpcarevise { However, ALCOP is the first to model how pipelining stage numbers affect performance and trade-offs between pipelining and tiling. ALCOP differs from analytical model-based search in that it combines ML- and analytical-based search as detailed in Section~\ref{sec:autotune:model-guided-tuning}.}

Recently, static analysis has arisen to supplement the standard ML-based schedule tuning, whose cost model lacks hardware knowledge. Tuna~\cite{wang2021tuna} builds a performance model for CPU and GPU to replace the ML-based schedule tuning in AutoTVM~\cite{chen2018learning}. We show that a combination of an analytical model and machine learning can achieve greater search efficiency than the Tuna technique.

%% file: 8_conclusion.tex





\section{Conclusion}
This paper addresses the important need for \textbf{automatic pipelining} in deep learning compilers. Due to the large tiling size required to mitigate bandwidth constraints, inter-tile parallelism is inadequate for achieving high utilization, and intra-tile pipelining becomes essential. We propose the first compiler solution that supports \textit{multi-stage, multi-level} pipelining. Through introducing automatic pipelining, our compiler can generate GPU programs with an average 1.23$\times$ and maximally 1.73$\times$ speedup over vanilla TVM~\cite{chen2018tvm}. Additionally, we develop an analytical performance model which significantly improves the search efficiency of the schedule tuning process.

%% file: main.bbl
\begin{thebibliography}{10}
\providecommand{\url}[1]{#1}
\csname url@samestyle\endcsname
\providecommand{\newblock}{\relax}
\providecommand{\bibinfo}[2]{#2}
\providecommand{\BIBentrySTDinterwordspacing}{\spaceskip=0pt\relax}
\providecommand{\BIBentryALTinterwordstretchfactor}{4}
\providecommand{\BIBentryALTinterwordspacing}{\spaceskip=\fontdimen2\font plus
\BIBentryALTinterwordstretchfactor\fontdimen3\font minus
  \fontdimen4\font\relax}
\providecommand{\BIBforeignlanguage}[2]{{%
\expandafter\ifx\csname l@#1\endcsname\relax
\typeout{** WARNING: IEEEtran.bst: No hyphenation pattern has been}%
\typeout{** loaded for the language `#1'. Using the pattern for}%
\typeout{** the default language instead.}%
\else
\language=\csname l@#1\endcsname
\fi
#2}}
\providecommand{\BIBdecl}{\relax}
\BIBdecl

\bibitem{he2016deep}
K.~He, X.~Zhang, S.~Ren, and J.~Sun, ``Deep residual learning for image
  recognition,'' in \emph{CVPR}, 2016, pp. 770--778.

\bibitem{devlin2018bert}
J.~Devlin, M.-W. Chang, K.~Lee, and K.~Toutanova, ``Bert: Pre-training of deep
  bidirectional transformers for language understanding,'' \emph{arXiv preprint
  arXiv:1810.04805}, 2018.

\bibitem{naumov2019deep}
M.~Naumov, D.~Mudigere, H.-J.~M. Shi, J.~Huang, N.~Sundaraman, J.~Park,
  X.~Wang, U.~Gupta, C.-J. Wu, A.~G. Azzolini \emph{et~al.}, ``Deep learning
  recommendation model for personalization and recommendation systems,''
  \emph{arXiv preprint arXiv:1906.00091}, 2019.

\bibitem{nvidiav100}
Nvidia, ``Nvidia tesla {V100} {GPU} architecture,'' \emph{Data Sheet}, pp.
  20--21, 2020.

\bibitem{nvidiaa100}
------, ``Nvidia {A100} tensor core {GPU},'' \emph{Data Sheet}, pp. 20--21,
  2020.

\bibitem{cutlass}
\BIBentryALTinterwordspacing
------, ``Nvidia cutlass release v2.7,'' 2021. [Online]. Available:
  \url{https://github.com/NVIDIA/cutlass}
\BIBentrySTDinterwordspacing

\bibitem{yan2020demystifying}
D.~Yan, W.~Wang, and X.~Chu, ``Demystifying tensor cores to optimize
  half-precision matrix multiply,'' in \emph{2020 IEEE International Parallel
  and Distributed Processing Symposium (IPDPS)}.\hskip 1em plus 0.5em minus
  0.4em\relax IEEE, 2020, pp. 634--643.

\bibitem{dakkak2019accelerating}
A.~Dakkak, C.~Li, J.~Xiong, I.~Gelado, and W.-m. Hwu, ``Accelerating reduction
  and scan using tensor core units,'' in \emph{Proceedings of the ACM
  International Conference on Supercomputing}, 2019, pp. 46--57.

\bibitem{feng2021egemm}
B.~Feng, Y.~Wang, G.~Chen, W.~Zhang, Y.~Xie, and Y.~Ding, ``Egemm-tc:
  Accelerating scientific computing on tensor cores with extended precision,''
  in \emph{Proceedings of the 26th ACM SIGPLAN Symposium on Principles and
  Practice of Parallel Programming}, 2021, pp. 278--291.

\bibitem{chen2018tvm}
T.~Chen, T.~Moreau, Z.~Jiang, L.~Zheng, E.~Yan, H.~Shen, M.~Cowan, L.~Wang,
  Y.~Hu, L.~Ceze \emph{et~al.}, ``{TVM}: An automated {End-to-End} optimizing
  compiler for deep learning,'' in \emph{13th USENIX Symposium on Operating
  Systems Design and Implementation (OSDI 18)}, 2018, pp. 578--594.

\bibitem{katel2022mlir}
N.~Katel, V.~Khandelwal, and U.~Bondhugula, ``Mlir-based code generation for
  gpu tensor cores,'' in \emph{Proceedings of the 31st ACM SIGPLAN
  International Conference on Compiler Construction}, 2022, pp. 117--128.

\bibitem{niu2021dnnfusion}
W.~Niu, J.~Guan, Y.~Wang, G.~Agrawal, and B.~Ren, ``Dnnfusion: accelerating
  deep neural networks execution with advanced operator fusion,'' in
  \emph{Proceedings of the 42nd ACM SIGPLAN International Conference on
  Programming Language Design and Implementation}, 2021, pp. 883--898.

\bibitem{zhao2022apollo}
J.~Zhao, X.~Gao, R.~Xia, Z.~Zhang, D.~Chen, L.~Chen, R.~Zhang, Z.~Geng,
  B.~Cheng, and X.~Jin, ``Apollo: Automatic partition-based operator fusion
  through layer by layer optimization,'' \emph{Proceedings of Machine Learning
  and Systems}, vol.~4, pp. 1--19, 2022.

\bibitem{asplos22-zhenzheng}
\BIBentryALTinterwordspacing
Z.~Zheng, X.~Yang, P.~Zhao, G.~Long, K.~Zhu, F.~Zhu, W.~Zhao, X.~Liu, J.~Yang,
  J.~Zhai, S.~L. Song, and W.~Lin, ``Astitch: Enabling a new multi-dimensional
  optimization space for memory-intensive ml training and inference on modern
  simt architectures,'' in \emph{Proceedings of the 27th ACM International
  Conference on Architectural Support for Programming Languages and Operating
  Systems}, ser. ASPLOS 2022.\hskip 1em plus 0.5em minus 0.4em\relax New York,
  NY, USA: Association for Computing Machinery, 2022, p. 359–373. [Online].
  Available: \url{https://doi.org/10.1145/3503222.3507723}
\BIBentrySTDinterwordspacing

\bibitem{zheng2020fusionstitching}
Z.~Zheng, P.~Zhao, G.~Long, F.~Zhu, K.~Zhu, W.~Zhao, L.~Diao, J.~Yang, and
  W.~Lin, ``Fusionstitching: boosting memory intensive computations for deep
  learning workloads,'' \emph{arXiv preprint arXiv:2009.10924}, 2020.

\bibitem{cublas}
\BIBentryALTinterwordspacing
NVIDIA, ``{cuBLAS},'' \url{https://docs.nvidia.com/cuda/cublas/index.html}.
  [Online]. Available: \url{https://docs.nvidia.com/cuda/cublas/index.html}
\BIBentrySTDinterwordspacing

\bibitem{xing2022bolt}
J.~Xing, L.~Wang, S.~Zhang, J.~Chen, A.~Chen, and Y.~Zhu, ``Bolt: Bridging the
  gap between auto-tuners and hardware-native performance,'' \emph{Proceedings
  of Machine Learning and Systems}, vol.~4, pp. 204--216, 2022.

\bibitem{xla}
\BIBentryALTinterwordspacing
Google, ``Xla: Optimizing compiler for machine learning,'' 2021. [Online].
  Available: \url{https://www.tensorflow.org/xla}
\BIBentrySTDinterwordspacing

\bibitem{chen2018learning}
T.~Chen, L.~Zheng, E.~Yan, Z.~Jiang, T.~Moreau, L.~Ceze, C.~Guestrin, and
  A.~Krishnamurthy, ``Learning to optimize tensor programs,'' \emph{arXiv
  preprint arXiv:1805.08166}, 2018.

\bibitem{lym2019delta}
S.~Lym, D.~Lee, M.~O'Connor, N.~Chatterjee, and M.~Erez, ``Delta: Gpu
  performance model for deep learning applications with in-depth memory system
  traffic analysis,'' in \emph{2019 IEEE International Symposium on Performance
  Analysis of Systems and Software (ISPASS)}.\hskip 1em plus 0.5em minus
  0.4em\relax IEEE, 2019, pp. 293--303.

\bibitem{radford2019language}
A.~Radford, J.~Wu, R.~Child, D.~Luan, D.~Amodei, I.~Sutskever \emph{et~al.},
  ``Language models are unsupervised multitask learners,'' \emph{OpenAI blog},
  vol.~1, no.~8, p.~9, 2019.

\bibitem{simonyan2014very}
K.~Simonyan and A.~Zisserman, ``Very deep convolutional networks for
  large-scale image recognition,'' \emph{arXiv preprint arXiv:1409.1556}, 2014.

\bibitem{cudnn}
\BIBentryALTinterwordspacing
NVIDIA, ``{cuDNN},''
  \url{https://docs.nvidia.com/deeplearning/cudnn/developer-guide/index.html}.
  [Online]. Available:
  \url{https://docs.nvidia.com/deeplearning/cudnn/developer-guide/index.html}
\BIBentrySTDinterwordspacing

\bibitem{Tavarageri2021}
S.~Tavarageri, A.~Heinecke, S.~Avancha, B.~Kaul, G.~Goyal, and R.~Upadrasta,
  ``{PolyDL: Polyhedral Optimizations for Creation of High-performance DL
  Primitives},'' \emph{ACM Transactions on Architecture and Code Optimization},
  vol.~18, no.~1, pp. 1--25, 2021.

\bibitem{chen2016xgboost}
T.~Chen and C.~Guestrin, ``Xgboost: A scalable tree boosting system,'' in
  \emph{Proceedings of the 22nd acm sigkdd international conference on
  knowledge discovery and data mining}, 2016, pp. 785--794.

\bibitem{huang2019gpipe}
Y.~Huang, Y.~Cheng, A.~Bapna, O.~Firat, D.~Chen, M.~Chen, H.~Lee, J.~Ngiam,
  Q.~V. Le, Y.~Wu \emph{et~al.}, ``Gpipe: Efficient training of giant neural
  networks using pipeline parallelism,'' \emph{Advances in neural information
  processing systems}, vol.~32, 2019.

\bibitem{narayanan2019pipedream}
D.~Narayanan, A.~Harlap, A.~Phanishayee, V.~Seshadri, N.~R. Devanur, G.~R.
  Ganger, P.~B. Gibbons, and M.~Zaharia, ``Pipedream: generalized pipeline
  parallelism for dnn training,'' in \emph{Proceedings of the 27th ACM
  Symposium on Operating Systems Principles}, 2019, pp. 1--15.

\bibitem{barham2022pathways}
P.~Barham, A.~Chowdhery, J.~Dean, S.~Ghemawat, S.~Hand, D.~Hurt, M.~Isard,
  H.~Lim, R.~Pang, S.~Roy \emph{et~al.}, ``Pathways: Asynchronous distributed
  dataflow for ml,'' \emph{arXiv preprint arXiv:2203.12533}, 2022.

\bibitem{zheng2022alpa}
L.~Zheng, Z.~Li, H.~Zhang, Y.~Zhuang, Z.~Chen, Y.~Huang, Y.~Wang, Y.~Xu,
  D.~Zhuo, E.~P. Xing \emph{et~al.}, ``Alpa: Automating inter-and
  $\{$Intra-Operator$\}$ parallelism for distributed deep learning,'' in
  \emph{16th USENIX Symposium on Operating Systems Design and Implementation
  (OSDI 22)}, 2022, pp. 559--578.

\bibitem{liu2016cambricon}
S.~Liu, Z.~Du, J.~Tao, D.~Han, T.~Luo, Y.~Xie, Y.~Chen, and T.~Chen,
  ``Cambricon: An instruction set architecture for neural networks,'' in
  \emph{2016 ACM/IEEE 43rd Annual International Symposium on Computer
  Architecture (ISCA)}.\hskip 1em plus 0.5em minus 0.4em\relax IEEE, 2016, pp.
  393--405.

\bibitem{jouppi2017datacenter}
N.~P. Jouppi, C.~Young, N.~Patil, D.~Patterson, G.~Agrawal, R.~Bajwa, S.~Bates,
  S.~Bhatia, N.~Boden, A.~Borchers \emph{et~al.}, ``In-datacenter performance
  analysis of a tensor processing unit,'' in \emph{Proceedings of the 44th
  annual international symposium on computer architecture}, 2017, pp. 1--12.

\bibitem{sohrabizadeh2020end}
A.~Sohrabizadeh, J.~Wang, and J.~Cong, ``End-to-end optimization of deep
  learning applications,'' in \emph{Proceedings of the 2020 ACM/SIGDA
  International Symposium on Field-Programmable Gate Arrays}, 2020, pp.
  133--139.

\bibitem{liao2021ascend}
H.~Liao, J.~Tu, J.~Xia, H.~Liu, X.~Zhou, H.~Yuan, and Y.~Hu, ``Ascend: a
  scalable and unified architecture for ubiquitous deep neural network
  computing: Industry track paper,'' in \emph{2021 IEEE International Symposium
  on High-Performance Computer Architecture (HPCA)}.\hskip 1em plus 0.5em minus
  0.4em\relax IEEE, 2021, pp. 789--801.

\bibitem{wei2017automated}
X.~Wei, C.~H. Yu, P.~Zhang, Y.~Chen, Y.~Wang, H.~Hu, Y.~Liang, and J.~Cong,
  ``Automated systolic array architecture synthesis for high throughput cnn
  inference on fpgas,'' in \emph{Proceedings of the 54th Annual Design
  Automation Conference 2017}, 2017, pp. 1--6.

\bibitem{lai2019heterocl}
Y.-H. Lai, Y.~Chi, Y.~Hu, J.~Wang, C.~H. Yu, Y.~Zhou, J.~Cong, and Z.~Zhang,
  ``Heterocl: A multi-paradigm programming infrastructure for software-defined
  reconfigurable computing,'' in \emph{Proceedings of the 2019 ACM/SIGDA
  International Symposium on Field-Programmable Gate Arrays}, 2019, pp.
  242--251.

\bibitem{wang2021autosa}
J.~Wang, L.~Guo, and J.~Cong, ``Autosa: A polyhedral compiler for
  high-performance systolic arrays on fpga,'' in \emph{The 2021 ACM/SIGDA
  International Symposium on Field-Programmable Gate Arrays}, 2021, pp.
  93--104.

\bibitem{parashar2019timeloop}
A.~Parashar, P.~Raina, Y.~S. Shao, Y.-H. Chen, V.~A. Ying, A.~Mukkara,
  R.~Venkatesan, B.~Khailany, S.~W. Keckler, and J.~Emer, ``Timeloop: A
  systematic approach to dnn accelerator evaluation,'' in \emph{2019 IEEE
  international symposium on performance analysis of systems and software
  (ISPASS)}.\hskip 1em plus 0.5em minus 0.4em\relax IEEE, 2019, pp. 304--315.

\bibitem{ning1993novel}
Q.~Ning and G.~R. Gao, ``A novel framework of register allocation for software
  pipelining,'' in \emph{Proceedings of the 20th ACM SIGPLAN-SIGACT symposium
  on Principles of programming languages}, 1993, pp. 29--42.

\bibitem{govindarajan1996framework}
R.~Govindarajan, E.~R. Altman, and G.~R. Gao, ``A framework for
  resource-constrained rate-optimal software pipelining,'' \emph{IEEE
  Transactions on Parallel and distributed systems}, vol.~7, no.~11, pp.
  1133--1149, 1996.

\bibitem{wei2012software}
H.~Wei, J.~Yu, H.~Yu, M.~Qin, and G.~R. Gao, ``Software pipelining for stream
  programs on resource constrained multicore architectures,'' \emph{IEEE
  Transactions on Parallel and Distributed Systems}, vol.~23, no.~12, pp.
  2338--2350, 2012.

\bibitem{hong2009analytical}
S.~Hong and H.~Kim, ``An analytical model for a gpu architecture with
  memory-level and thread-level parallelism awareness,'' in \emph{Proceedings
  of the 36th annual international symposium on Computer architecture}, 2009,
  pp. 152--163.

\bibitem{volkov2016understanding}
V.~Volkov, \emph{Understanding latency hiding on GPUs}.\hskip 1em plus 0.5em
  minus 0.4em\relax University of California, Berkeley, 2016.

\bibitem{wang2020mdm}
L.~Wang, M.~Jahre, A.~Adileho, and L.~Eeckhout, ``Mdm: The gpu memory
  divergence model,'' in \emph{2020 53rd Annual IEEE/ACM International
  Symposium on Microarchitecture (MICRO)}.\hskip 1em plus 0.5em minus
  0.4em\relax IEEE, 2020, pp. 1009--1021.

\bibitem{huang2014gpumech}
J.-C. Huang, J.~H. Lee, H.~Kim, and H.-H.~S. Lee, ``Gpumech: Gpu performance
  modeling technique based on interval analysis,'' in \emph{2014 47th Annual
  IEEE/ACM International Symposium on Microarchitecture}.\hskip 1em plus 0.5em
  minus 0.4em\relax IEEE, 2014, pp. 268--279.

\bibitem{zhang2011quantitative}
Y.~Zhang and J.~D. Owens, ``A quantitative performance analysis model for gpu
  architectures,'' in \emph{2011 IEEE 17th international symposium on high
  performance computer architecture}.\hskip 1em plus 0.5em minus 0.4em\relax
  IEEE, 2011, pp. 382--393.

\bibitem{baghsorkhi2010adaptive}
S.~S. Baghsorkhi, M.~Delahaye, S.~J. Patel, W.~D. Gropp, and W.-m.~W. Hwu, ``An
  adaptive performance modeling tool for gpu architectures,'' in
  \emph{Proceedings of the 15th ACM SIGPLAN symposium on Principles and
  practice of parallel programming}, 2010, pp. 105--114.

\bibitem{wang2021tuna}
Y.~Wang, X.~Zhou, Y.~Wang, R.~Li, Y.~Wu, and V.~Sharma, ``Tuna: A static
  analysis approach to optimizing deep neural networks,'' \emph{arXiv preprint
  arXiv:2104.14641}, 2021.

\end{thebibliography}
